\documentclass[aps,prl,reprint,groupedaddress,
citeautoscript,
amsmath,amssymb,longbibliography,linenumbers,
]{revtex4-2}

\usepackage{graphicx}
\usepackage{amsmath}
\usepackage{siunitx}
\usepackage{xr-hyper} 
\usepackage[hidelinks]{hyperref} 
\usepackage[table]{xcolor}
\usepackage{booktabs}  
\usepackage[caption=false]{subfig} 
\usepackage{makecell,multirow,bigdelim}  
\usepackage{rotating,bm}   
\usepackage{mathptmx}  
\usepackage{enumitem}
\usepackage[normalem]{ulem}
\usepackage{xspace}
\xspaceaddexceptions{+}
\usepackage{float}
\setlist{nosep} 
\newcommand*{\citen}[1]{\cite{#1}}
\usepackage{verbatim}

\graphicspath{{figures-tables/}}

\setlist[description]{leftmargin=0pt,labelindent=0pt}
\DeclareMathAlphabet\mathbfcal{OMS}{cmsy}{b}{n}

\newcommand{\rtwoscan}{r\textsuperscript{2}SCAN\xspace}

\newcommand{\rot}[1]{\rotatebox{90}{#1}}
\newcommand{\mrx}{\multirow{2}{*}{x}}

\newcommand{\trid}{$P6_3/mmc$-tridymite\xspace}

\newcolumntype{P}[1]{>{\centering\arraybackslash}p{#1}}
\newcolumntype{M}[1]{>{\centering\arraybackslash}m{#1}}

\makeatletter
\newcommand*{\addFileDependency}[1]{
  \typeout{(#1)}
  \@addtofilelist{#1}
  \IfFileExists{#1}{}{\typeout{No file #1.}}
}
\makeatother

\newcommand*{\myexternaldocument}[1]{%
    \externaldocument[][nocite]{#1}%
    \addFileDependency{#1.tex}%
    \addFileDependency{#1.aux}%
}

\usepackage{anyfontsize} 

\usepackage{tikz}
\usetikzlibrary{decorations.pathreplacing}

\usepackage[figure,figure*]{hypcap} 
\usepackage[right,pagewise,modulo]{lineno} 
\usepackage{pdfpages} 
\usepackage{pgffor} 

\makeatletter
\AtBeginDocument{\let\LS@rot\@undefined}
\makeatother

\def\supplementfilename{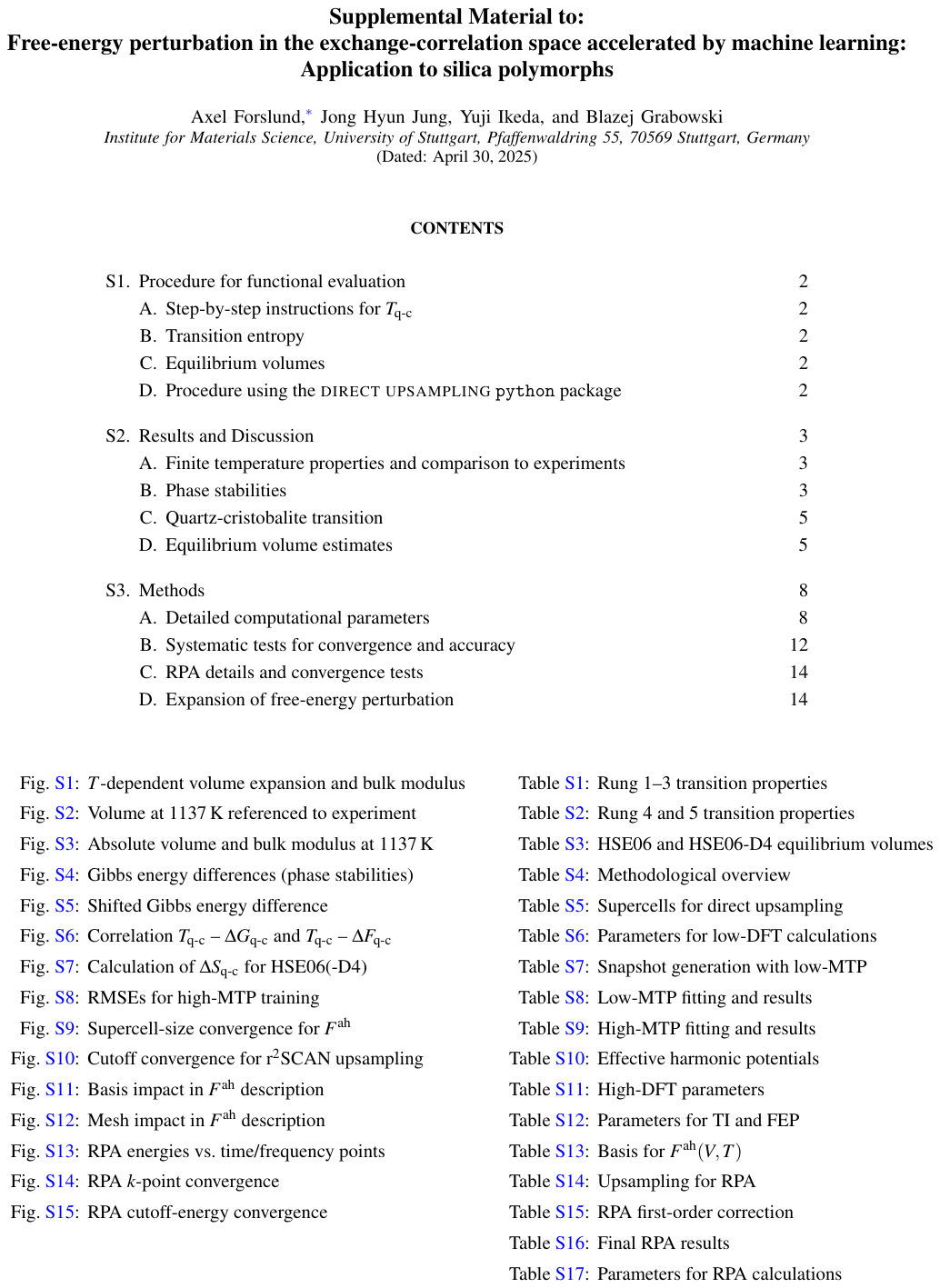}

\pdfximage{\supplementfilename}
\def\numbersupplementpages{\the\pdflastximagepages}

\newif\ifarXiv
\arXivtrue 

\begin{document}
\myexternaldocument{supplementary}

\title{Free-energy perturbation in the exchange-correlation space accelerated by machine learning: Application to silica polymorphs}
\author{Axel Forslund}
\email{axel.forslund@imw.uni-stuttgart.de}
\author{Jong Hyun Jung}
\author{Yuji Ikeda}
\author{Blazej Grabowski}
\affiliation{Institute for Materials Science, University of Stuttgart, Pfaffenwaldring 55, 70569 Stuttgart}
\date{\today}
\begin{abstract}
We propose a free-energy-perturbation approach accelerated by machine-learning potentials to efficiently compute transition temperatures and entropies for all rungs of Jacob's ladder. We apply the approach to the dynamically stabilized phases of SiO$_2$, which are characterized by challengingly small transition entropies. All investigated functionals from rungs 1--4 fail to predict an accurate transition temperature by 25--200\%. Only by ascending to the fifth rung, within the random phase approximation, an accurate prediction is possible, giving a relative error of 5\%. We provide a clear-cut procedure and relevant data to the community for, e.g., developing and evaluating new functionals.
\end{abstract}
\maketitle
%
 
The search for the ultimate exchange-correlation functional in density functional theory (DFT) is the most significant challenge for first-principles calculations in materials science. The flora of approximations was divided into rungs on Jacob's ladder by Perdew and Schmidt~\cite{perdew01,perdew05}. Starting from the local density and generalized gradient approximation (LDA and GGA), the sophistication increases up to the highest chemical accuracy. At the top resides the random phase approximation (RPA), which considers exact exchange together with an approximate many-body treatment of correlations~\cite{langreth77,kaltak14}. Being nonlocal, the latter also effectively includes dispersion interactions, while, for lower rungs, semiempirical dispersion corrections may be required. Despite the increased sophistication, the higher-level approximations do not always lead to improved results~\cite{schimka13,xiao13,tran16}. Evaluation against experiments is thus critical and has, conventionally, been performed for 0-K properties such as lattice constants and cohesive energies~\cite{staroverov03,haas09b,schimka11,cohen12,schimka13,xiao13,sun15a,tran16,zhang18c,kingsbury22,kothakonda23}, see Table~\ref{tab:literature-functionals}. In a few cases, low-temperature approximations were used~\cite{wei24}, which, however, can lead to errors at elevated temperatures~\cite{grabowski19}. Even worse, dynamically unstable phases, such as titanium alloys, high entropy alloys, certain perovskites, and silica, cannot be treated at all.

\begin{table}[b]
\caption{\textit{Ab initio} studies evaluating the performance of the exchange-correlation treatment (mGGA=meta-GGA, hGGA=hybrid GGA; 0\,K equilibrium properties: $a_0$=lattice constant, $B$=bulk modulus, $E_\textrm{coh}$=cohesion energy, $E_\textrm{form}$=formation energy; finite temperature ($T$) properties: $T_\textrm{q-c}$=transition temperature between $\beta$-quartz and $\beta$-cristobalite in SiO$_2$, $\Delta S_{\textrm{q-c}}$=corresponding transition entropy).
}
\label{tab:literature-functionals}
\centering
\small
\begin{ruledtabular} \begin{tabular}{c@{}l ll c@{}c@{}c@{}c@{}c c}
    &&& &\rot{LDA}&\rot{GGA}&\rot{mGGA}&\rot{hGGA}&\rot{RPA}&\\
    Year & Ref. & Properties& \multicolumn{1}{r}{Rung$\rightarrow$}&1&2&3&4&5& No.\ of solids\\
    \midrule
    2003&\citen{tao03}&$a_0$, $B$& \rdelim\}{12}{*}[\,0~K] &x&x&x&&& 18\\
    2009&\citen{perdew09}&$a_0$, $E_\textrm{coh}$& &x&x&x&&& 36\\
    2009&\citen{haas09b}&$a_0$, $B$& &x&x&x&&& 60\\
    2011&\citen{schimka11}&$a_0$, $E_\textrm{coh/form}$& &&x&&x&x& 30\\
    2011&\citen{sun11}&$a_0$, $E_\textrm{coh}$& &x&x&x&&& 20\\
    2013&\citen{schimka13}&$a_0$, $E_\textrm{coh}$& &&x&x&&x& 30\\
    2013&\citen{xiao13}&$a_0$, $B$&  &x&x&x&x&x& 6\\
    2015&\citen{sun15a}&$a_0$& &x&x&x&&& 20\\
    2016&\citen{tran16}&$a_0$, $E_\textrm{coh}$, $B$& &x&x&x&x&& 49\\
    2018&\citen{zhang18c}&$a_0$, $E_\textrm{form}$& &&x&x&&& 200\\
    2022&\citen{kingsbury22}&$a_0$, $E_\textrm{form}$& &&x&x&&& 6000\\
    2023&\citen{kothakonda23}&$a_0$, $E_\textrm{form}$& &&x&x&&& 1000\\
    \midrule
    \multicolumn{2}{c}{\multirow{2}{*}{This work}} & \multicolumn{2}{l}{\multirow{2}{2.9cm}{finite $T$, $T_\textrm{q-c}$, $\Delta S_{\textrm{q-c}}$, dynamically stabilized}} &\mrx&\mrx&\mrx&\mrx&\mrx& \multirow{2}{1.4cm}{SiO$_2$ $\beta$-phases}\\
    \\
\end{tabular} \end{ruledtabular}
\end{table}

Silica, SiO$_2$, is often used as a model system by virtue of its technologically relevant polymorphs~\cite{demuth99,xiao13}. At higher temperatures, the $\beta$-phases are stable, i.e., $\beta$-quartz, $\beta$-cristobalite, and $\beta$-tridymite (\trid)~\cite{schnurre04}. These phases are dynamically stabilized and, thus, require explicit finite-temperature modeling. An additional challenge is the stringent requirement on the statistical and computational precision of the Gibbs energy differences when targeting phase transitions. The Gibbs energies of the $\beta$-phases are similar in magnitude and temperature dependence, so small entropy differences drive the transitions. Even minute inaccuracies of a few meV/atom can shift transition temperatures by several hundred K.

Here, we develop an approach that facilitates the prediction of high-accuracy transition temperatures and entropies for all rungs of Jacob's ladder. A key component is free-energy perturbation in the exchange-correlation space. The approach works genuinely at finite temperatures, enabling the treatment of dynamically stabilized phases with the same high precision as phases that are stable already at 0\,K. We apply the approach to the $\beta$-phases of silica, focusing on the transition between $\beta$-quartz and $\beta$-cristobalite. We compute the transition temperature $T_{\textrm{q-c}}$ and transition entropy $\Delta S_{\textrm{q-c}}$ between these two phases for all rungs of Jacob's ladder up to the RPA level. Analysis of the results provides a simple geometrical understanding of the phase-space differences between the different exchange-correlation treatments.

\begin{figure*}[tbh]
    \includegraphics{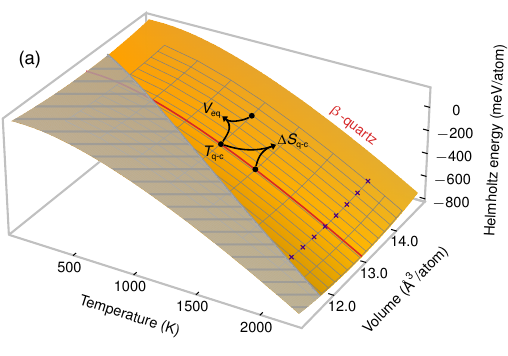}
    \centering \hfill
    \includegraphics{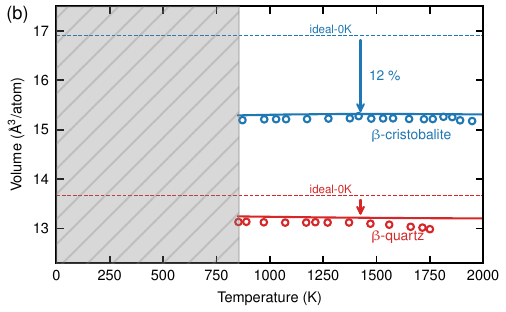}
    \caption{(a) Helmholtz energy surface for $\beta$-quartz within LDA. The red line marks the volume expansion at ambient pressure. The black dots indicate volume-temperature points for the calculation of the transition temperature, $T_\textrm{q-c}$, transition entropy, $\Delta S_\textrm{q-c}$, and equilibrium volume, $V_\textrm{eq}$, for rungs 4 and 5 within the proposed approach. The purple crosses indicate the points used for training the machine-learning potential. (b)~Volume expansion at ambient pressure of $\beta$-quartz and $\beta$-cristobalite within \rtwoscan-D4. Finite-temperature results (solid lines) are compared with experimental data (circles)~\cite{berger66,ackermann74,wright75,ohsumi84,swainson95,bourova98} and results for symmetry-constrained idealized structures at 0\,K (dashed lines). The gray areas mark the dynamical instability regime. See Fig.~S1 \cite{supplement} for volume-expansion and bulk-modulus results for various rung 1-3 functionals.
    }
    \label{fig:stability_regime}
\end{figure*}

The proposed approach is based on the direct upsampling method \cite{jung23}. Within the latter, machine-learning interatomic potentials are utilized to efficiently compute a high-precision Helmholtz energy surface with well-converged computational parameters (e.g., a dense volume-temperature mesh and supercells with several thousand atoms), isolating the inherent error due to the exchange-correlation approximation and enabling its evaluation. Direct upsampling can be applied well for rungs 1--3 from Jacob's ladder. For the present case, we compute Helmholtz energy surfaces for the silica $\beta$-phases utilizing various functionals from rungs 1--3 (LDA~\cite{ceperley80}, GGA-PBE~\cite{perdew96}, meta-GGA \rtwoscan~\cite{furness20} with and without dispersion corrections). The $\beta$-quartz LDA surface is shown in Fig.~\ref{fig:stability_regime}(a) as an example. The gray-shaded region marks the instability regime where the system transforms into $\alpha$-quartz, highlighting the necessity of an explicit finite-temperature treatment. Various thermodynamic quantities can be extracted from the Helmholtz energy surface, e.g., the thermal expansion, as emphasized by the red line.
Figure~\ref{fig:stability_regime}(b) compares the expansion for $\beta$-quartz and $\beta$-cristobalite within \rtwoscan, including the D4 correction with experiments marked by circles. The explicit finite-temperature results (solid lines) give a good prediction, while calculations for the symmetry-constrained idealized 0-K structures are meaningless (dashed lines). A comparison of various functionals from rungs 1--3 with experiments is provided in Sec.~S2~\cite{supplement}. LDA shows a good agreement for the volume at $T_\textrm{q-c}$ with a deviation below 1.5\%, but fails to correctly predict the temperature dependence. The third rung meta-GGA functionals, including dispersion corrections, give the best results with deviations below 1\%, qualitatively reproducing the temperature dependence.

\begin{figure}[t]
    \centering
    \includegraphics{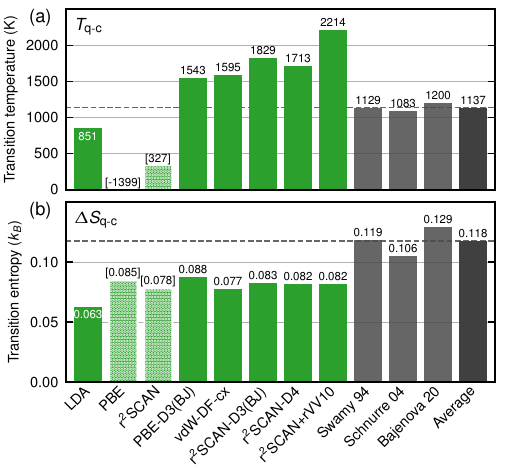}
    \caption{%
    $\beta$-quartz to $\beta$-cristobalite (a) transition temperatures $T_{\textrm{q-c}}$ and (b) transition entropies $\Delta S_{\textrm{q-c}}$ in SiO$_2$ for rungs 1--3 functionals (green) compared with CALPHAD~\cite{swamy94,schnurre04,bajenova20} (gray). Extrapolated temperatures are indicated with brackets, and the CALPHAD average is marked with a dashed line.
    }
    \label{fig:transition}
\end{figure}

The crossing of the $\beta$-quartz and $\beta$-cristobalite Gibbs energies determines $T_{\textrm{q-c}}$. The Gibbs energies are obtained by a Legendre transformation of the Helmholtz energy surfaces. The resulting $T_{\textrm{q-c}}$'s from rungs 1--3 are compared in Fig.~\ref{fig:transition}(a) with CALPHAD values representing experiments. The spread of predicted values is significant. Without dispersion corrections, $T_{\textrm{q-c}}$ is underestimated. LDA's $T_{\textrm{q-c}}$ is 251\,K (22\%) below CALPHAD, while PBE's and \rtwoscan's predictions even fall out of the stability window of the phases. Including dispersion corrections shifts $T_{\textrm{q-c}}$ upward, resulting in overestimations, e.g. 408\,K (36\%) for PBE-D3(BJ) or 594\,K (52\%) for \rtwoscan-D4. The corresponding transition entropies $\Delta S_{\textrm{q-c}}$ (differences between the slopes of the Gibbs energies at $T_{\textrm{q-c}}$) are shown in Fig.~\ref{fig:transition}(b). All rung 1--3 functionals underestimate $\Delta S_{\textrm{q-c}}$ with the largest discrepancy of 56\% for LDA.

In light of the discrepancies in predicting $T_{\textrm{q-c}}$ and $\Delta S_{\textrm{q-c}}$ observed for the functionals of rungs 1--3, an extension of the analysis to higher rungs is desirable. The difficulty is that the computational requirements for rungs 4 and 5 increase strongly. In the following, we describe a systematically extendable approach that overcomes these difficulties and enables---in successive steps---an accurate prediction of the transition temperature and entropy for rungs 4 and 5.

The approach builds on two main observations extracted from the wide range of results obtained for rungs 1--3. (1)~The transition temperature $T_{\textrm{q-c}}$ is well correlated with $\Delta F_{\textrm{q-c}}$, the Helmholtz energy difference between the two phases for a certain, specified condition. (2)~Free energy perturbation can be utilized to efficiently obtain Helmholtz energy differences between the functionals. Using these observations, it is possible to drastically reduce the computational effort in predicting $T_{\textrm{q-c}}$ and $\Delta S_{\textrm{q-c}}$ within rungs 4 and~5. Observation~(1) allows us to focus on a single volume-temperature point for each of the two phases for the $T_{\textrm{q-c}}$ prediction (cf.~Fig.~\ref{fig:stability_regime}(a) for $\beta$-quartz), reducing the number of calculations by about two orders of magnitude. The entropy $\Delta S_{\textrm{q-c}}$ can be obtained by one additional point on each surface. Observation (2) dispenses with the necessity to fit machine-learning potentials for rungs 4 or~5. Instead, the optimized potentials from rungs 1-3 are used to generate suitable snapshots. Taken together, the prediction of $T_{\textrm{q-c}}$ and $\Delta S_{\textrm{q-c}}$ for rungs 4 or~5 can be obtained in a few hundred snapshots of supercells with about two hundred atoms.

\begin{figure}[t]
    \centering
    \includegraphics{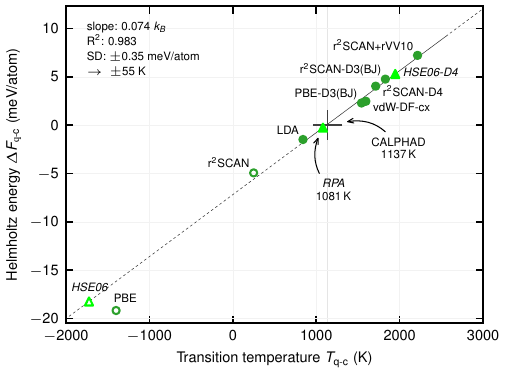}
    \caption{Correlation between the predicted transition temperature and Helmholtz energy difference between $\beta$-quartz and $\beta$-cristobalite at $T^\text{exp}_\text{q-c}$, and $V_\text{cristobalite}^\text{exp}$ and $V_\text{quartz}^\text{exp}$. The filled dark green circles show explicitly computed results and the line a linear fit of them. The hollow circles mark extrapolations out of the stability regime of the phases. The light green triangles mark predictions for rung 4 and 5 functionals (labeled in Italics) utilizing the fit.}
    \label{fig:correlation}
\end{figure}

The correlation between $T_{\textrm{q-c}}$ and $\Delta F_\textrm{q-c}$ is displayed in Fig.~\ref{fig:correlation}. The difference $\Delta F_\textrm{q-c}$ ($y$-axis) is obtained as
\begin{align}
\label{eq:deltaF}
    \Delta F_\textrm{q-c} &= F_\textrm{cristobalite}(V_\textrm{cristobalite}^\textrm{exp},T^\textrm{exp}_\textrm{q-c}) - F_\textrm{quartz}(V_\textrm{quartz}^\textrm{exp},T^\textrm{exp}_\textrm{q-c}),
\end{align}
where $F_\textrm{cristobalite}$ and $F_\textrm{quartz}$ correspond to the Helmholtz energies for a given exchange-correlation functional; further, $T^\textrm{exp}_\textrm{q-c}$ is the experimental (or CALPHAD) transition temperature, and $V_\textrm{cristobalite}^\textrm{exp}$ and $V_\textrm{quartz}^\textrm{exp}$ are the experimental equilibrium volumes of the two phases at $T^\textrm{exp}_\textrm{q-c}$. The $T_{\textrm{q-c}}$ value ($x$-axis) corresponds to the self-consistently computed transition temperature of the respective functional. The such obtained $(T_{\textrm{q-c}},\Delta F_\textrm{q-c})$-pairs are shown by the dark green dots in Fig.~\ref{fig:correlation}. The good linear relation is quantified by the small standard deviation of the fit of 0.35~meV/atom, which propagates into an uncertainty of 55~K in the transition temperature prediction. Even the predictions from \rtwoscan and PBE, which fall outside the stability regime of the phases, are reasonably captured by extrapolation. This extrapolative capacity is useful for locating functionals with low predictability of $T_\textrm{q-c}$, e.g., HSE06 (see Fig.~\ref{fig:correlation}). However, for the main purpose of locating functionals with high predictability, we rely only on the interpolative behavior of the linear relation. The best prediction is quantified by a vanishing $\Delta F_\textrm{q-c}$, which corresponds to the averaged CALPHAD transition temperature of 1137\,K.

To obtain the transition entropy $\Delta S_{\textrm{q-c}}$, we utilize a finite difference,
\begin{align}
\label{eq:entropy}
    \Delta S_{\textrm{q-c}} = (\Delta F_\textrm{q-c} - \Delta F'_\textrm{q-c}) /  \Delta T,
\end{align}
where $\Delta F'_\textrm{q-c}$ is a second Helmholtz energy difference computed similarly as $\Delta F_\textrm{q-c}$ in Eq.~\eqref{eq:deltaF} but at a shifted temperature. The linear temperature dependence of the free energy difference between the two phases (Figs.~S4, S5, S7 \cite{supplement}) allows us to utilize a large enough temperature shift $\Delta T$ ($\approx300$\,K) to guarantee a numerically stable evaluation of $\Delta S_{\textrm{q-c}}$.

\begin{figure*}[bt]
    \centering
    \includegraphics[]{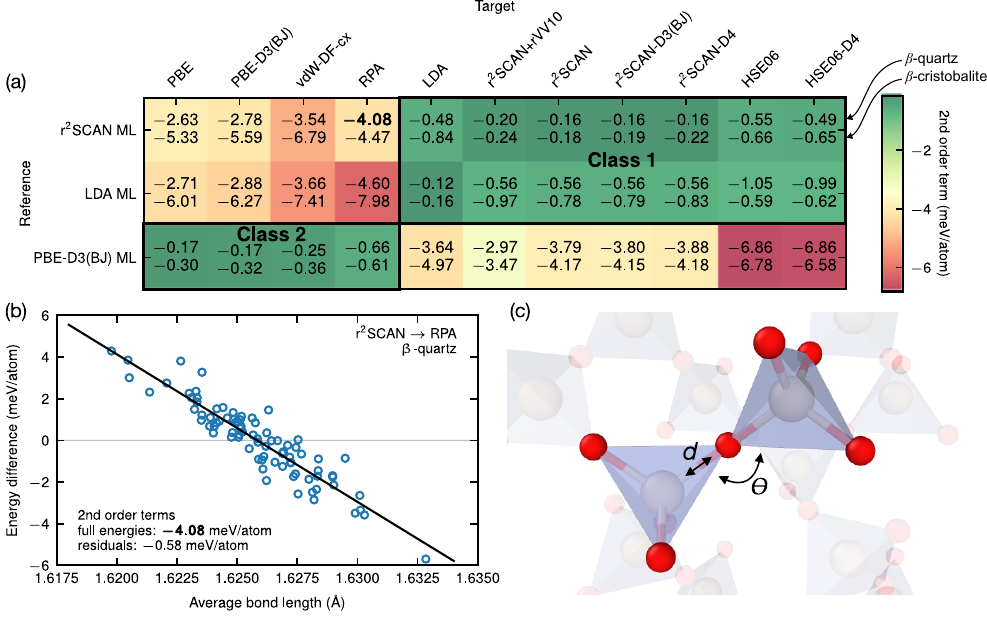}
    \caption{(a) Second-order terms for $\beta$-quartz / $\beta$-cristobalite, in meV/atom. The fields are colored according to the average value and qualitatively indicate the resemblance between the phase spaces of the reference and target functionals.
    (b) Correlation between average Si--O bond length and the energy difference between \rtwoscan and RPA for $\beta$-quartz (blue circles). The black line shows a linear fit. Second-order terms from free-energy perturbation using full energy differences and residuals are written in the lower left corner. The values on the $y$-axis are offset with respect to the average energy. (c) Illustration of the Si--O bond length $d$ and Si--O--Si bond angle $\theta$ in silica.}
    \label{fig:2nd_order_term__correlation__tetrahedron}
\end{figure*}

To compute the Helmholtz energy for each of the phases in Eq.~\eqref{eq:deltaF} for functionals of rung 4 or 5, we utilize free-energy perturbation in the exchange-correlation space (second observation). Specifically, at volume $V$ and temperature $T$,
\begin{align}
\label{eq:F_ML_up}
    F(V,T) = F^\textrm{ML}(V,T; \textrm{xc}^\textrm{1--3}) + \Delta F^{\textrm{up}}(V,T; \textrm{xc}^\textrm{1--3}).
\end{align}
Here, $F^\textrm{ML}(V,T; \textrm{xc}^\textrm{1--3})$ is the Helmholtz energy obtained within direct upsampling with a machine-learning (ML) potential fitted to an exchange-correlation (xc) functional from rung 1--3 and
\begin{align}
\label{eq:perturbation}
      \Delta F^{\textrm{up}} (V,T; \textrm{xc}^\textrm{1--3})  = - k_B T \ln{\left< \exp{\left(-\frac{\Delta E}{k_B T} \right)} \right>_\textrm{ML}},
\end{align}
where $\Delta E = E-E^\textrm{ML}$, with the energies $E$ and $E^\textrm{ML}$ calculated with a rung 4 or 5 functional and the $\textrm{xc}^\textrm{1--3}$ ML potential, respectively. The $\textrm{xc}^\textrm{1--3}$ ML potential is also used to sample the thermodynamic average in Eq.~\eqref{eq:perturbation}.

In practice, the convergence behavior of the thermodynamic average with the number of snapshots is decisive. To quantify the convergence behavior, we make use of the second-order terms in the expansion of Eq.~\eqref{eq:perturbation} (see Sec.~S3~D \cite{supplement}). Generally, values less than 1~meV/atom allow for efficient use of Eq.~\eqref{eq:perturbation} (around 100 snapshots for high convergence). Values for various combinations of functionals are given in the matrix in Fig.~\ref{fig:2nd_order_term__correlation__tetrahedron}(a). Two classes of functionals can be distinguished as highlighted by the two green areas identifying small second-order terms and thus a good overlap between the corresponding functionals in phase space. The first, larger class is composed of LDA, meta-GGA (\rtwoscan), and the hybrid functional HSE06 with or without dispersion corrections. The second class consists of the GGA functionals (PBE with or without correction), vdW-DF-cx (classified as a nonlocal van der Waals functional \cite{berland14}), and RPA.

Interclass combinations of functionals show larger second-order terms (yellow to red areas in the matrix). The larger second-order terms can be understood by simple geometrical means. In Fig.~\ref{fig:2nd_order_term__correlation__tetrahedron}(b), we plot, as an example, the energy differences between \rtwoscan (represented by the machine-learning potential) and RPA as a function of the average Si--O bond length [Fig.~\ref{fig:2nd_order_term__correlation__tetrahedron}(c)]. The correlation is good and, in particular, removing the spread in energies due to the bond-length dependence leads to a second-order term of below 1 meV/atom (from $-4.08$ to $-0.58$~meV/atom). Si--O--Si bond angles are similarly well correlated as bond lengths---however, the combined account for bond lengths and angles does not improve the correlation, which shows that these two parameters are strongly correlated.
The importance of bond lengths and angles has been realized since the early days of the development of classical potentials for SiO$_2$~\cite{vashishta90}.
Here, we unveil how even fine nuances in the description of these coordinates (changes in the range of 0.01\AA) affect the free-energy perturbation.

We utilize the proposed approach to compute $T_{\textrm{q-c}}$ and $\Delta S_{\textrm{q-c}}$ for HSE06, HSE06-D4 (both rung 4), and RPA (rung 5) by upsampling, respectively, from \rtwoscan for the first two and from PBE-D3(BJ) for RPA, chosen according to the smallest corresponding second-order term. The results are included in Fig.~\ref{fig:correlation} by the bright green triangles. For the hybrid-GGA HSE06 functional, the extrapolated transition temperature lies far off, close to that of PBE. The D4 corrections strongly shift the transition temperature upward, resulting in an overestimation. The transition entropy is 0.060 $k_B$ and 0.084 $k_B$, respectively. Overall, no improvement is seen compared to the rung 1--3 functionals. For RPA, our analysis (Sec.~S3~C \cite{supplement}) shows that extremely well-converged computational settings are required, in particular for the plane wave cutoff (1000\,eV), to obtain convergence. We also find that core polarization is important, shifting $T_{\textrm{q-c}}$ up by 267 K.
The final RPA transition temperature and entropy are 1081~K and 0.088 $k_B$ (underestimation of 5\% and 25\%), thus being closest to experiment among the investigated exchange-correlation treatments.

The proposed approach is systematically extendable and---based on the provided data \cite{axel25data} and clear-cut procedure, Sec.~S1 \cite{supplement}---straightforwardly available to the community. For example, in developing new functionals, $T_{\textrm{q-c}}$ can be used as the first benchmark quantity. The set of dedicated snapshots (100 for each surface) along with the ML energies for the various rung 1--3 functionals is available for download \cite{axel25data}. To evaluate a new functional, the energies for the snapshots are computed (with any independent code being possible) and utilized in the free-energy perturbation formulas. The predictive capability of the functional can also be tested for the transition entropy by utilizing another set of snapshots. Further, we provide a third set of snapshots with which the equilibrium volume of the new functional at $T_{\textrm{q-c}}$ can be predicted [cf.~Fig.~\ref{fig:stability_regime}(a) and Secs.~S1~C, S2~D~\cite{supplement}]. Eventually, at a larger computational cost, the full Helmholtz energy surface can be upsampled, giving access to various thermodynamic quantities (e.g., heat capacity, bulk modulus). Importantly, with the full surface available, the predicted quantities no longer rely on the correlation between $T_{\textrm{q-c}}$ and $\Delta F_{\textrm{q-c}}$ nor the linearity of the free energy differences.

The investigated SiO$_2$ system entails various challenges (dynamical instability, small transition entropies, soft phonon modes \cite{tezuka91}). The successful application of the introduced approach to SiO$_2$ indicates that other material systems should be likewise treatable, thus opening the door to a new area of functional development and evaluation at finite temperatures.

\begin{acknowledgments}
\textit{Acknowledgements.}
We appreciate fruitful discussions with Nikolay Zotov. 
This project has received funding from the European Research Council (ERC) under the European Union's Horizon 2020 research and innovation program (grant agreement No 865855).
This work was also funded by the German Research Foundation (DFG), Germany---Project-ID 358283783-SFB 1333/2 2022.
The authors acknowledge support by the state of Baden-Württemberg through bwHPC and the German Research Foundation (DFG) through grant No.~INST 40/575-1 FUGG (JUSTUS 2 cluster).
The authors gratefully acknowledge the scientific support and HPC resources provided by the Erlangen National High-Performance Computing Center (NHR@FAU) of the Friedrich-Alexander-Universität Erlangen-Nürnberg (FAU) under the NHR project a102cb. NHR funding is provided by federal and Bavarian state authorities. NHR@FAU hardware is partially funded by the German Research Foundation (DFG)---440719683.
B.G. acknowledges the support by the Stuttgart Center for Simulation Science (SimTech).
Y.I.~is funded by the DFG---519607530.
\end{acknowledgments}



%

\ifarXiv
    

\section*{Supplementary material}

## S1. PROCEDURE FOR FUNCTIONAL EVALUATION

The transition temperature, $T_{\text{q-c}}$, and the transition entropy, $\Delta S_{\text{q-c}}$, between $\beta$-quartz and $\beta$-cristobalite can be estimated for any new functional and/or DFT setup using the approach established in the main text, together with snapshots provided online [1] (in `input_functional_evaluation`).

### A. Step-by-step instructions for $T_{\text{q-c}}$

**Step 1**: Snapshots for the $T_{\text{q-c}}$ calculation are downloaded (subfolder `a.transition_temperature`; file format: VASP POSCAR). They are uncorrelated (separated by enough steps) and have been pre-generated for both phases (`quartz_beta`, `cristobalite_beta`) with three machine-learning (ML) potentials (`lda_mtp`, `pbe_d3bj_mtp`, `r2scan_mtp`).

**Step 2**: The energies of the snapshots are calculated by the user with the new functional and/or DFT setup.

**Step 3**: The second-order term (Sec. S3 D) is determined for all potentials and both phases as

$$e_2 = -\frac{1}{2k_B T_{\text{q-c}}^{\text{exp}}} \left[ \frac{1}{N_{\text{s}}} \sum_{i=1}^{N_{\text{s}}} \Delta E_i^2 - \left( \frac{1}{N_{\text{s}}} \sum_{i=1}^{N_{\text{s}}} \Delta E_i \right)^2 \right], \quad \text{(S1)}$$

with $T_{\text{q-c}}^{\text{exp}} = 1137$ K and $\Delta E_i = E_i - E_i^{\text{ML}}$, the difference between the user-calculated and ML-potential energy for the $i$'th snapshot with $N_{\text{a}}$ atoms. The sums run over the provided $N_{\text{s}} = 100$ snapshots for each phase. The ML energies, $E_i^{\text{ML}}$, are provided in the file `energy_mtp` in the snapshot folder. The potential giving the smallest $e_2$'s is selected for the next steps. The $e_2$'s should be $\lesssim 2$ meV/atom to guarantee convergence within the number of provided snapshots. Larger values indicate the inability to predict $T_{\text{q-c}}$.

**Step 4**: The Helmholtz energy difference between the new functional/DFT setup and the selected ML potential is determined with free-energy perturbation. In particular, for each phase ($p$ = cristobalite or quartz),

$$\Delta F_p^{\text{up}} = -\frac{k_B T_{\text{q-c}}^{\text{exp}}}{N_{\text{a}}} \ln \left[ \frac{1}{N_{\text{s}}} \sum_{i=1}^{N_{\text{s}}} \exp \left( -\frac{\Delta E_i}{k_B T_{\text{q-c}}^{\text{exp}}} \right) \right]. \quad \text{(S2)}$$

**Step 5**: The Helmholtz energy difference between $\beta$-quartz and $\beta$-cristobalite is calculated by

$$\Delta F_{\text{q-c}} = \Delta F_{\text{cristobalite}}^{\text{up}} - \Delta F_{\text{quartz}}^{\text{up}} + \Delta F_{\text{q-c}}^{\text{ML}}, \quad \text{(S3)}$$

where the first two terms are given by Eq. (S2), and $\Delta F_{\text{q-c}}^{\text{ML}}$ is the Helmholtz energy difference on the level of the ML potential, evaluated in this work and given as (1137 K-column for $T_{\text{q-c}}$; unit: meV/atom):

|  | 1137 K | 1419 K |
|---|---|---|
| LDA MTP: | $-1.34$ | $-2.84$ |
| PBE-D3(BJ) MTP: | $+2.07$ | $+0.12$ |
| r$^2$SCAN MTP: | $-5.30$ | $-6.90$ |

**Step 6**: Finally, the transition temperature is calculated by

$$T_{\text{q-c}} \approx T_{\text{q-c}}^{\text{exp}} + \frac{\Delta F_{\text{q-c}}}{0.074 \, k_B} \quad \text{(S4)}$$

and compared with the CALPHAD average value $T_{\text{q-c}}^{\text{exp}} = 1137$ K. Equation (S4) follows from the relation of $T_{\text{q-c}}$ with the Helmholtz energy difference $\Delta F_{\text{q-c}}$ between the two phases, as established in the main text (see also Sec. S2 C). The value $0.074 \, k_B$ corresponds to the slope of this relation.

### B. Transition entropy

The transition entropy, $\Delta S_{\text{q-c}}$, can be computed by repeating steps 1, 2, 4, 5 for a new set of snapshots (subfolder `b.transition_entropy`) for the selected ML potential. These snapshots correspond to $T = 1419$ K, which should be used in Eq. (S2). Also, the 1419 K-column should be used in Eq. (S3). The second Helmholtz energy difference $\Delta F_{\text{q-c}}'$ obtained in this way is used together with $\Delta F_{\text{q-c}}$ (i.e., the difference obtained during the $T_{\text{q-c}}$ calculation) in Eq. (2) of the main text with $\Delta T = 282$ K to obtain the transition entropy.

### C. Equilibrium volumes

The equilibrium volumes of $\beta$-quartz and $\beta$-cristobalite can be likewise estimated for the new functional/DFT setup by repeating steps 1, 2, and 4 for a third set of snapshots (subfolder `c.equilibrium_volume`). These snapshots correspond to the experimental transition temperature as for the $T_{\text{q-c}}$ calculation, but are located at different volumes. Using these snapshots in Eq. (S2) gives new $\Delta F_p^{\text{up}\prime\prime}$ values which, together with the original $\Delta F_p^{\text{up}}$ values from the $T_{\text{q-c}}$ prediction, give correction slopes,

$$v_p = (\Delta F_p^{\text{up}\prime\prime} - \Delta F_p^{\text{up}})/\Delta V, \quad \text{(S5)}$$

where $\Delta V = 0.8$ Å$^3$/atom. The linear corrections given by the $v_p$'s need to be added to the original volume-dependent Helmholtz energies for the selected ML potentials for each phase (the offsets do not affect the equilibrium volume and are not relevant). The original Helmholtz energies are stored in text format in the sub-subfolder `helmholtz_energies`. The equilibrium volumes of the new functional/DFT setup can be obtained by finding the minima of the new curves.

### D. Procedure using the DIRECT UPSAMPLING python package

The procedure of preparing input files and post-processing (steps 2–6) is automated using scripts provided in Ref. [1]. The scripts utilize the DIRECT UPSAMPLING python package, available through https://gitlab.com/axefor/direct-upsampling, which requires the Atomic Simulation Environment (ASE) python library [2] for input file generation. This allows any compatible calculator class to be used, for example, that of VASP as in the provided example.

## S2. RESULTS AND DISCUSSION

We have computed Helmholtz energy surfaces for four different phases of silica ($SiO_2$); specifically, for

1. $\beta$-quartz [3]
2. $\beta$-cristobalite [4, 5]
3. $P6_3/mmc$-tridymite [6]
4. $C222_1$-tridymite [7]

For each phase, eight different exchange-correlation functionals from rungs 1–3 of Jacob's ladder, without and with dispersion corrections, were used (see also Sec. S3 A):

1. LDA
2. PBE
3. r$^2$SCAN
4. PBE-D3(BJ)
5. vdW-DF-cx
6. r$^2$SCAN-D3(BJ) ⎫
7. r$^2$SCAN-D4     ⎬ dispersion corrected
8. r$^2$SCAN+rVV10  ⎭

In total, 32 densely sampled Helmholtz energy surfaces were obtained. In addition, following the approach proposed in the main text, three volume-temperature $(V, T)$ points were computed with the rung-4 hybrid functional HSE06 with and without the D4 dispersion correction for both the $\beta$-quartz and $\beta$-cristobalite phases, in order to predict $T_{\text{q-c}}$, $\Delta S_{\text{q-c}}$, and the equilibrium volume of each phase. For the rung-5 random phase approximation (RPA), two $(V, T)$ points were computed for each of the two phases to predict $T_{\text{q-c}}$ and $\Delta S_{\text{q-c}}$.

In the following, we show data complementary to the main text, including temperature dependencies of the equilibrium volumes and bulk moduli, as well as phase stabilities, for the rung-1–3 functionals. We also provide tables including the data plotted in the main text for the $\beta$-quartz–$\beta$-cristobalite transition, as well as some additional results obtained with the developed approach.

### A. Finite temperature properties and comparison to experiments

Complementary to Fig. 1(b), we show in Fig. S1(a) and (c) detailed results for the volume expansion of $\beta$-quartz and $\beta$-cristobalite for all the functionals in rungs 1–3. Additionally, in Fig. S1(b) and (d), we show the corresponding adiabatic bulk moduli. The results are compared with experimental data, where available.

A further comparison with experiments, with the temperature fixed to 1137 K, is shown in Fig. S2. The PBE functional significantly overestimates the volume for all phases, while LDA performs quite well. Interestingly, the van der Waals corrected PBE functional, PBE-D3(BJ), gives a similar volume as r$^2$SCAN, while r$^2$SCAN with a van der Waals

correction gives the best results. The difference between different types of van der Waals corrections is small in this case (whether local or nonlocal), although r$^2$SCAN-D3(BJ) seems to give the smallest discrepancy with experiments in terms of volume.

We also show bar plots for the absolute volume and the adiabatic bulk modulus at the same temperature of 1137 K in Fig. S3, including additionally $C222_1$-tridymite. The relation between the volumes of the different phases is consistently predicted by all functionals. Turning to the bulk moduli, $\beta$-quartz is predicted to be the hardest, consistent with previous predictions [15]. In terms of bulk modulus, the r$^2$SCAN functional, with or without dispersion corrections, gives the results closest to experiments where available (only for $\beta$-quartz).

### B. Phase stabilities

Figure S4 shows the stability of $\beta$-cristobalite and the two tridymite phases with respect to $\beta$-quartz at ambient pressure. The Gibbs energy differences between the phases are below 10 meV/atom over almost the whole investigated temperature range for most functionals. Only for the PBE functional, there is an overall larger difference, destabilizing $\beta$-quartz with respect to the other phases. This prediction, that $\beta$-quartz would be metastable, stands in sharp contrast to the experimental phase diagram. The r$^2$SCAN functional produces qualitatively similar incorrect results as PBE, but with smaller energy differences.

The introduction of van der Waals corrections stabilizes $\beta$-quartz in all cases. The relative stability between the two tridymites and $\beta$-cristobalite is less affected. This is correlated with the difference in density between the phases: because quartz is denser than the other phases, it gets a stronger stabilizing contribution from the dispersion interactions. The tridymites and $\beta$-cristobalite, on the other hand, have densities very similar to each other.

The $P6_3/mmc$-tridymite and $\beta$-cristobalite phases have energies that are extremely close to each other, on a level smaller than the precision of the present calculations. It is therefore not possible in the present work to distinguish when or if there is a transition in pure $SiO_2$. In previous work [17–20], however, it has been suggested that $P6_3/mmc$-tridymite is not formed in pure $SiO_2$, but only stabilized by impurities. The reason for this could be kinetics or a small energy difference that is not resolvable by DFT calculations. In the present study, we rely on the previous assessments [17, 18] and focus on the $\beta$-quartz–$\beta$-cristobalite transition, which is well and consistently described within CALPHAD.

We note, though, that all the rung-1–3 functionals predict that the $P6_3/mmc$-tridymite phase is more stable than the $C222_1$-tridymite phase within the investigated temperature range. This is consistent with experiments [6], which have suggested a transition temperature of 693 K.

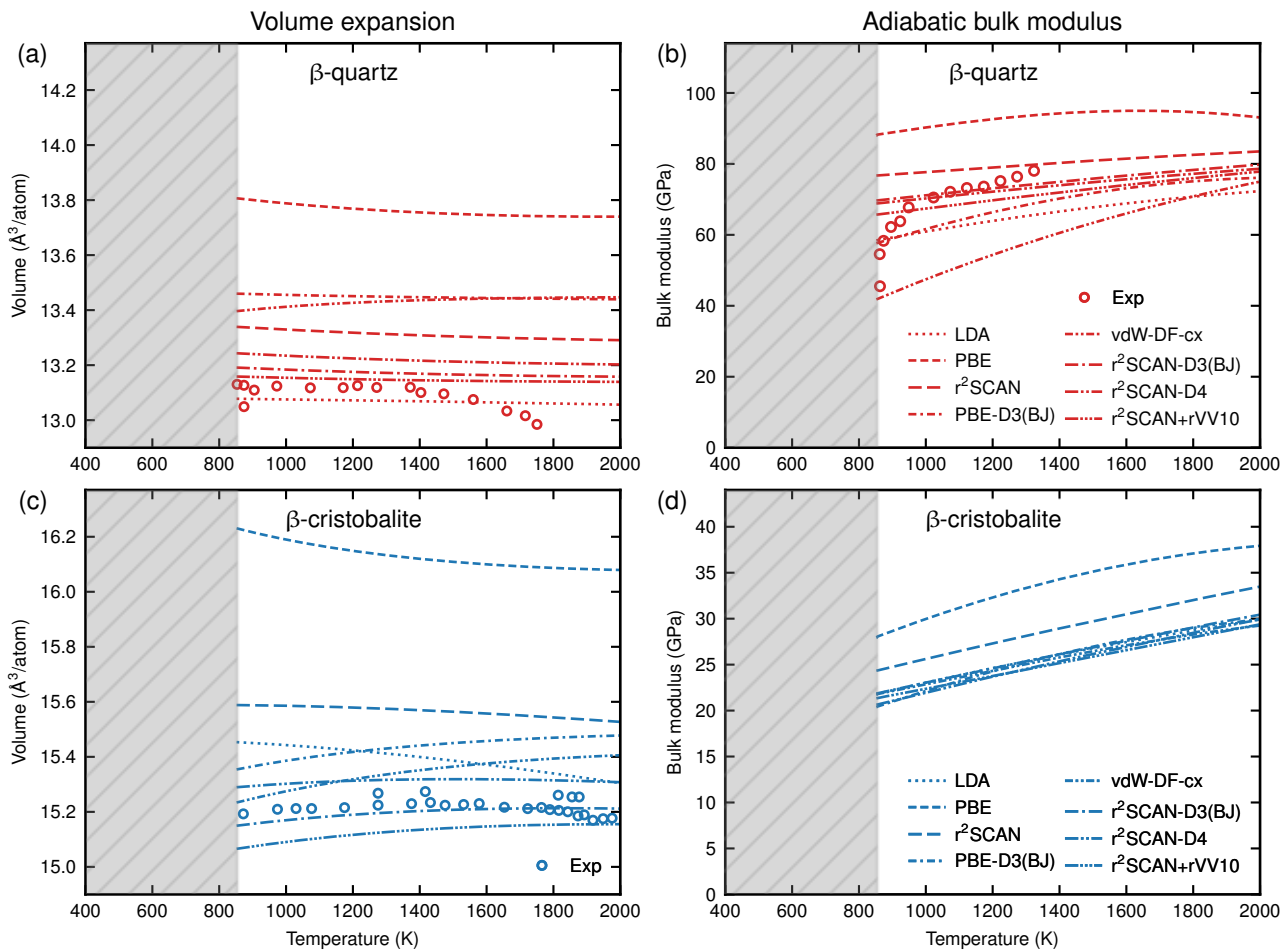

FIG. S1. Rung-1–3 finite-temperature volume [(a) and (c)] and adiabatic bulk modulus [(b) and (d)], for $\beta$-quartz [(a) and (b)] and $\beta$-cristobalite [(c) and (d)]. The experimental data (circles) are taken from the following references: $\beta$-quartz [8–10], $\beta$-cristobalite [9, 11, 12], and Refs. [13, 14] as assessed in Ref. [9].

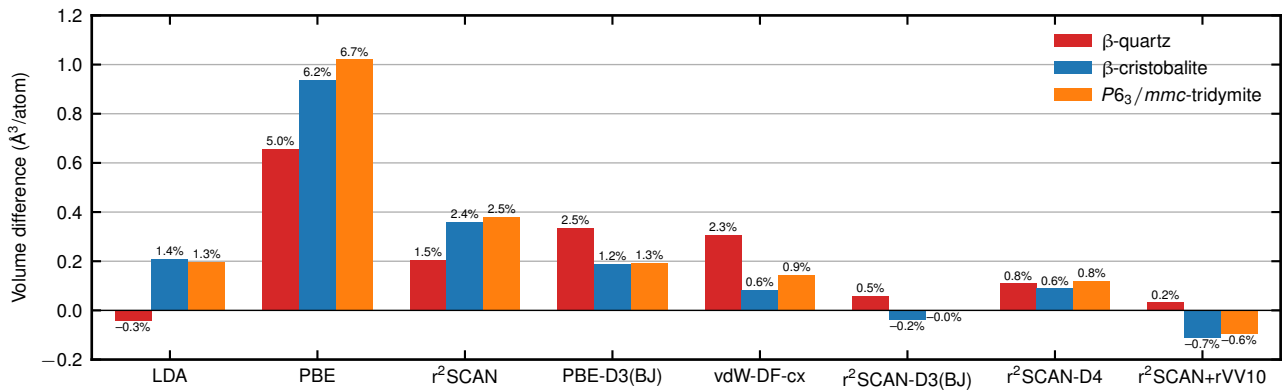

FIG. S2. Rung-1–3 volume difference with respect to experiments [8, 9] at 1137 K of $\beta$-quartz, $\beta$-cristobalite and $P6_3/mmc$-tridymite, respectively. The volumes are referenced with respect to a fit to experimental data of the respective structure. The numbers over the bars indicate the relative difference in percent.

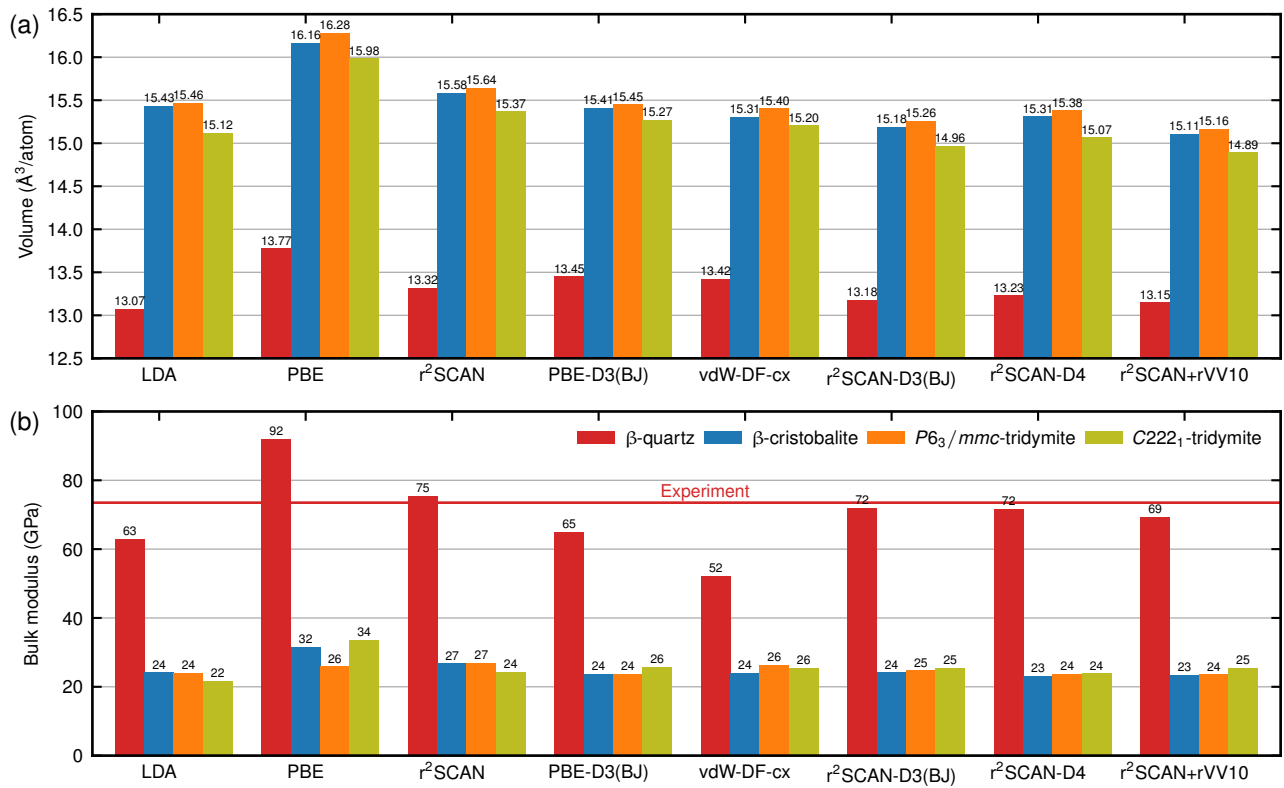

FIG. S3. Rung-1–3 absolute (a) volume and (b) adiabatic bulk modulus at 1137 K of $\beta$-quartz, $\beta$-cristobalite and $P6_3/mmc$- and $C222_1$-tridymite. The red line marks the experimental bulk modulus for quartz [10, 16].

### C. Quartz-cristobalite transition

Transition properties between $\beta$-quartz and $\beta$-cristobalite derived from the Gibbs energies of the rung-1–3 functionals (Fig. S4) are shown in Table S1. To estimate transition properties for functionals of rungs 4 and 5, we utilize the observed similarity of the $\Delta G(T)$ dependencies for the lower rungs. To see this similarity of the $\Delta G(T)$'s between $\beta$-cristobalite and $\beta$-quartz clearly, we plot *shifted* Gibbs energy differences in Fig. S5. The shifts have been adjusted to make all curves cross at 1137 K, i.e., at the experimental transition temperature. All investigated rung 1–3 functionals show very linear temperature dependencies with similar slopes. The corresponding transition entropies differ by only a few $0.01 \, k_B$.

Because of the similar $\Delta G(T)$'s between $\beta$-cristobalite and $\beta$-quartz, the $T_{q\text{-}c}$'s obtained for the different functionals are well correlated with corresponding Gibbs energy differences. We plot this correlation for the functionals of rungs 1–3 in Fig. S6(a). The difference in Gibbs energy is calculated at 1137 K and ambient pressure by optimizing the volumes. For comparison, the correlation of the $T_{q\text{-}c}$'s with the respective Helmholtz energy differences (1137 K, volume fixed to 13.2/15.2 Å$^3$/atom for $\beta$-quartz/$\beta$-cristobalite) for the functionals of rungs 1–3 (corresponding to Fig. 3 in the main text) are repeated in Fig. S6(b). The correlation is very strong in

both cases, resulting in a slope of $0.074 \, k_B$. The $R^2$ value is slightly better for the Gibbs energy, corresponding to a standard deviation of the fit of 0.1 meV/atom or 16 K, as compared to 0.35 meV/atom or 55 K for the Helmholtz energy. The correlation shown in Fig. S6(b), jointly with the free-energy perturbation in the exchange-correlation space, is the basis of the efficient $T_{q\text{-}c}$ prediction for higher-rung functionals in the proposed approach. Predicted $T_{q\text{-}c}$'s are given in Table S2.

The linearity of the $\Delta G(T)$'s between $\beta$-cristobalite and $\beta$-quartz also enables an efficient and numerically stable prediction of $\Delta S_{q\text{-}c}$. Specifically, a linear correction to the slopes of the Helmholtz energy with respect to temperature is computed [see Eq. (2) in the main text]. For HSE06 and HSE06-D4, the Helmholtz energy difference was computed at three temperature points (1135, 1419 and 1702 K) to validate the assumption of linearity. The correction is indeed very linear as seen in Fig. S7. For RPA, two points were used, at 1135 and 1419 K. The resulting transition entropies are given in Table S2.

### D. Equilibrium volume estimates

The equilibrium volume of a certain phase at a given temperature can be estimated for a higher-rung functional by up-sampling from a rung-1–3 functional at two volume points. This is possible due to the linear (in fact, nearly constant)

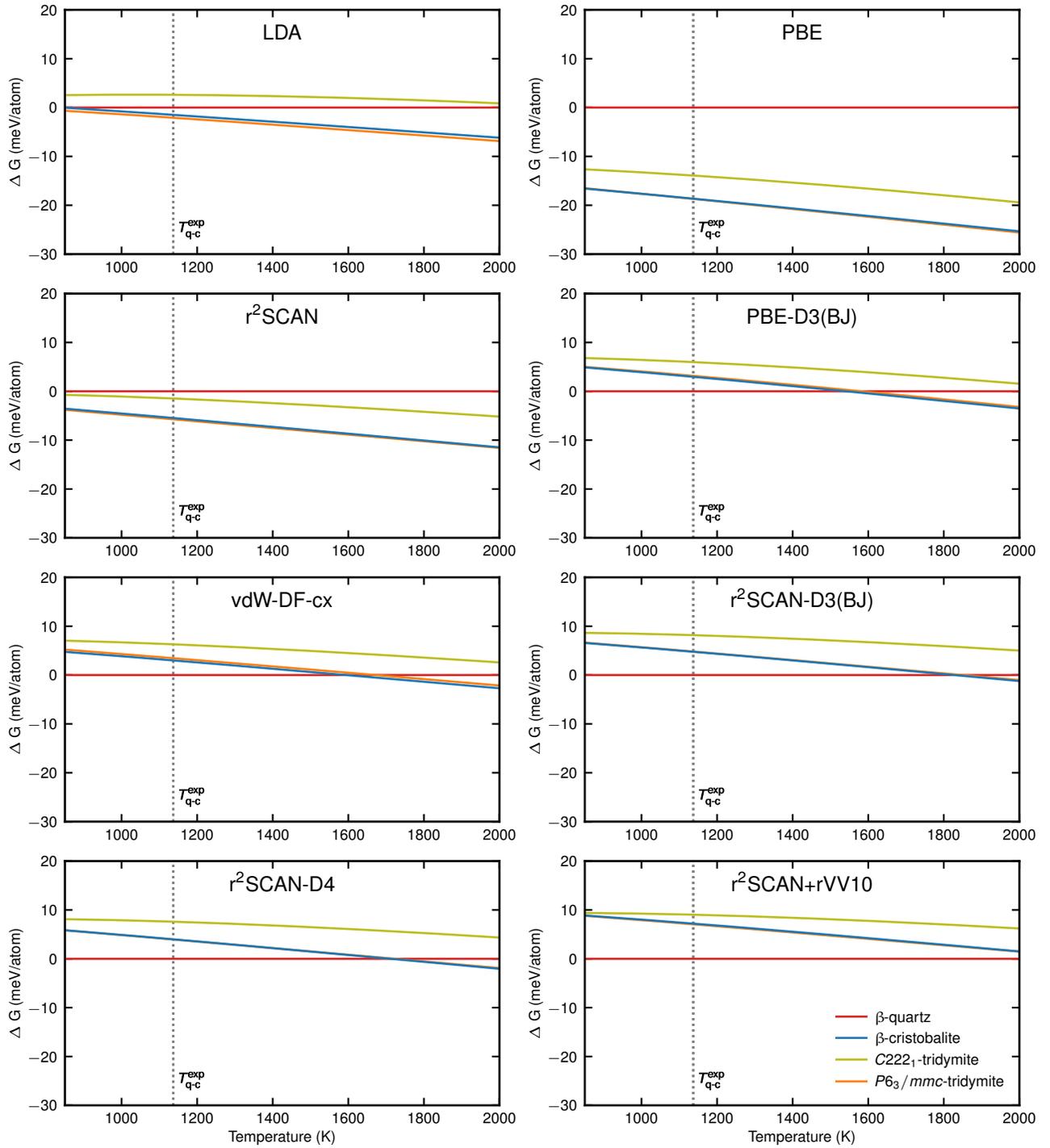

FIG. S4. Gibbs energy difference $\Delta G(T)$ of $\beta$-cristobalite and $C222_1$- and $P6_3/mmc$-tridymite with respect to $\beta$-quartz at ambient pressure for the functionals in rungs 1–3. If not visible, the orange line is below the blue one. The vertical dashed line marks the CALPHAD transition temperature $T_{\text{q-c}}^{\text{exp}} = 1137$ K.

volume dependence of the upsampling term. Utilizing the r²SCAN MTP as the reference, we have estimated the equilibrium volumes for $\beta$-quartz and $\beta$-cristobalite for the rung-4 hybrid HSE06 and HSE06-D4 functionals at 1137 K. The upsampling was performed at 13.2 and 14 Å³/atom, and 15.2 and 16 Å³/atom, for $\beta$-quartz and $\beta$-cristobalite, respectively. The estimated equilibrium volumes are given in Table S3. These equilibrium volumes and the impact of the dispersion correction are consistent with the rung-1–3 functionals, in particular with PBE/PBE-D3(BJ), for $\beta$-quartz and $\beta$-cristobalite.

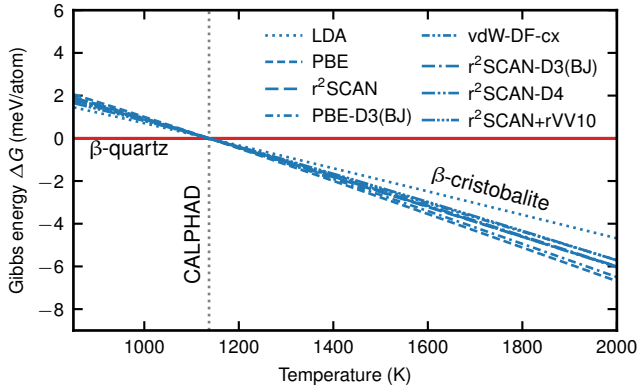

FIG. S5. Shifted Gibbs energy of $\beta$-cristobalite with respect to $\beta$-quartz, at ambient pressure. The curves have been shifted to be aligned at the CALPHAD transition temperature of 1137 K.

TABLE S1. Rung 1–3 transition properties.

|  | $T_{\text{q-c}}$ (K) | Deviation (%) | $\Delta G_{\text{q-c}}$ (meV/atom) | $\Delta S_{\text{q-c}}$ ($k_B$) |
|---|---|---|---|---|
| LDA | 886 | −22 | −1.21 | 0.052 |
| PBE[a] | −1399 | −223 | −18.65 | 0.085 |
| r$^2$SCAN[a] | 243 | −79 | −5.36 | 0.067 |
| PBE-D3(BJ) | 1545 | 36 | 2.98 | 0.088 |
| vdW-DF-cx | 1595 | 40 | 3.00 | 0.077 |
| r$^2$SCAN-D3(BJ) | 1843 | 62 | 4.90 | 0.083 |
| r$^2$SCAN-D4 | 1731 | 52 | 4.06 | 0.081 |
| r$^2$SCAN+rVV10 | 2229 | 96 | 7.20 | 0.082 |
| CALPHAD Average | 1137 | 0 | 0.00 | 0.118 |

[a] The values corresponding to these functionals are extrapolated from the dynamically stabilized regime.

TABLE S2. Rung 4 and 5 transition properties.

|  | $T_{\text{q-c}}$ (K) | Deviation (%) | $\Delta F_{\text{q-c}}$ (meV/atom) | $\Delta S_{\text{q-c}}$ ($k_B$) |
|---|---|---|---|---|
| HSE06[a] | −1723 | −252 | −18.25 | 0.060 |
| HSE06-D4 | 1945 | 71 | 5.30 | 0.084 |
| RPA | 1081 | −5 | −0.26 | 0.088 |
| CALPHAD Average | 1137 | 0 | 0.00 | 0.118 |

[a] The value corresponding to HSE06 is extrapolated.

TABLE S3. Rung 4 equilibrium volumes (Å$^3$/atom) at $T_{\text{q-c}}^{\text{exp}}$.

|  | $\beta$-quartz | $\beta$-cristobalite |
|---|---|---|
| HSE06 | 13.48 | 16.35 |
| HSE06-D4 | 13.22 | 15.29 |

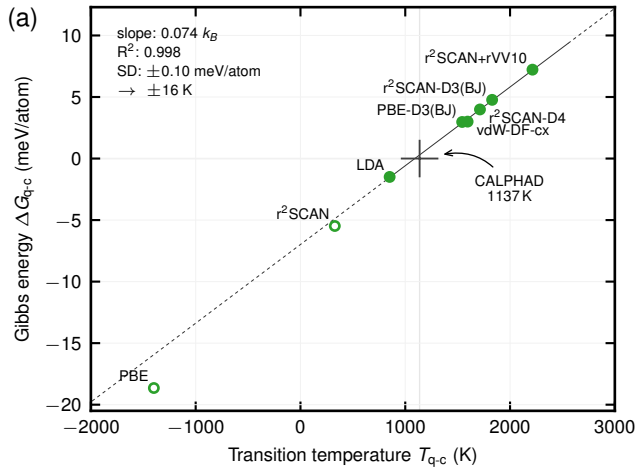

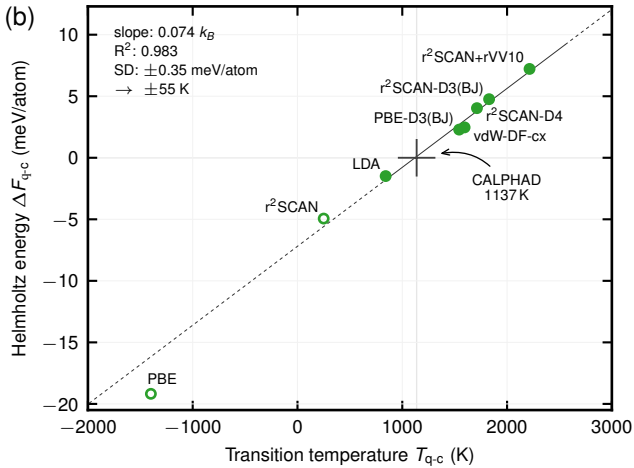

FIG. S6. Correlation between the predicted transition temperature and the (a) Gibbs energy difference $\Delta G_{\text{q-c}}$ and (b) Helmholtz energy difference between $\beta$-quartz and $\beta$-cristobalite at $T_{\text{q-c}}^{\text{exp}}$, in (b) with fixed volumes. The filled dark green circles show explicitly computed results and the line a linear fit of them. The hollow circles mark extrapolations out of the stability regime of the phases. (b) is a repetition for rungs 1–3 from Fig. 3 in the main text.

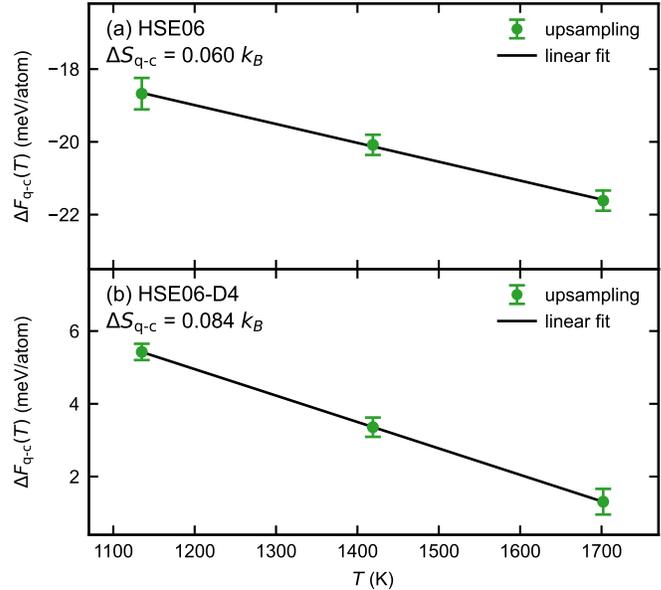

FIG. S7. Calculation of the transition entropy, $\Delta S_{\text{q-c}}$, from the slopes of the Helmholtz energy, $\Delta F_{\text{q-c}}$, for (a) HSE06 and (b) HSE06-D4. The error bars denote statistical errors within a 95% confidence interval.

## S3. METHODS

A condensed overview reflecting the computational complexity of the different steps/methods used in this work can be found in Table S4. For a detailed step-by-step description and analysis of the direct-upsampling method, see Ref. [21]. In the following sections, we present only the relevant methodological details and parameters used in the present work, as well as specific convergence tests and RPA parameters. In addition, we also supply the expansion for free-energy perturbation.

### A. Detailed computational parameters

The DFT calculations were performed with the projector-augmented wave (PAW) method [22, 23] as implemented in the Vienna *ab initio* simulation package (VASP) [24]. For all calculations for rungs 1–4, the PAW potentials with the valence-electron configuration Si: $3s^2 3p^2$; O: $2s^2 2p^4$ were used. For the RPA calculations, the VASP _GW PAW potentials were used, as well as an additional PAW potential for the Si core-polarization correction, see Sec. S3 C. The utilized functionals from rungs 1–3 follow:

| | |
|---|---|
| LDA | local density approximation [25] |
| GGA-PBE | generalized gradient approximation parametrized by Perdew, Burke, Ernzerhof [26] |
| $r^2$SCAN | regularized-restored strongly constrained and appropriately normed meta-GGA functional [27] |
| vdW-DF-cx | nonlocal van der Waals density functional with consistent exchange [28] |
| GGA-PBE-D3(BJ) | [29] |
| $r^2$SCAN-D3(BJ) | [29, 30] |
| $r^2$SCAN-D4 | [30] |
| $r^2$SCAN+rVV10 | [31] |

where

| | |
|---|---|
| D3(BJ) | van der Waals interaction of Grimme with Becke–Johnson damping [29] |
| D4 | van der Waals interaction of Grimme [32] |
| rVV10 | revised nonlocal van der Waals functional [33] |

From rungs 4 and 5, we used:

| | |
|---|---|
| HSE06 | Heyd–Scuseria–Ernzerhof hybrid functional [34] |
| HSE06-D4 | [32, 35] |
| RPA | random phase approximation within the adiabatic connection fluctuation-dissipation theorem [36, 37] |

The direct upsampling was performed in the extended version of Ref. [38], due to the dynamic instability at 0 K of the investigated phases. In particular, effective equations of state were used as the basis for the finite temperature calculations. These equations of state were adjusted to allow for reasonable thermodynamic integration. Further, an effective harmonic reference fitted to high temperatures was used as a reference for the thermodynamic integration. The supercell sizes for the different steps of the direct upsampling method are summarized in Table S5.

In Table S6, the parameters for the low-DFT molecular dynamics (MD) runs for the initial low-MTP training are given. Table S7 contains parameters for the snapshot generation with the low-MTP. The parameters for the low and high-MTP are given in Tables S8 and Table S9, respectively. The training energy and force root-mean-square errors (RMSEs) of different levels for the high-MTP are shown in Fig. S8(a) and (b), respectively. The corresponding RMSEs for a test set (of equal number of atoms to the training set) are shown in Fig. S8(c) and (d).

The parameters for the effective harmonic potentials fitted to high-temperature snapshots calculated with high-DFT are compiled in Table S10. Effective *harmonic* potentials were used instead of quasiharmonic ones because of a non-trivial volume dependence of the anharmonic contribution, mainly due to soft modes corresponding to the rotation of $SiO_4$ tetrahedra. Further, effective harmonic potentials were only fitted for LDA and then used as a reference for all functionals.

The parameters for the high-DFT calculations of the snapshots are given in Table S11. The same DFT settings apply to the upsampling. The utilized $k$-point densities from 192 to 288 $k$ points·atoms (kp·atoms) are sufficient for high convergence due to the insulating character of $SiO_2$, as explicitly confirmed (see Sec. S3 B). This contrasts with metallic systems, where at least an order of magnitude denser $k$-point samplings are required (e.g., Ref. [38]). Regarding the occupation function, all calculations were performed by applying the Gaussian smearing with a width of 0.1 eV.

Thermodynamic-integration and upsampling grids and supercell sizes can be seen in Table S12. The anharmonic Helmholtz energy $F^{ah}$—explicitly available at specific $V$ and $T$ points from the previous steps—was parametrized in $V$ and $T$ with the basis sets shown in Table S13.

For $\beta$-quartz, a fixed $c/a$ lattice-parameter ratio was used for all simulations. It was fixed to 1.1, which is close to the experimental value [39]. In addition, careful tests were performed at $c/a$-ratios of 1.09 and 1.08, and the effect on Helmholtz energy was below 0.5 meV/atom even for these relatively large changes of the ratio. In experiments and calculations with LDA and GGA for $\alpha$-quartz and idealized $\beta$-quartz at 0 K, the changes in $c/a$ ratio were found to be much smaller [15]. These small changes are expected to have a negligible impact on the Helmholtz energy. For $P6_3/mmc$-tridymite, a fixed $c/a$ ratio of 1.64 was used.

TABLE S4. Methodological overview. Steps 1 and 2 are part of the "standard" direct-upsampling method, including both thermodynamic integration (TI) and free-energy perturbation (FEP). Steps 3 and 4 are new developments in this work. Details regarding the fourth step are given in Sec. S3 C.

| Step | Method | Level | Supercell | No. atoms | No. $V, T$ points | No. MD steps/snapshots | Plane-wave cutoff (eV) |
|---|---|---|---|---|---|---|---|
| 1 | TI | MTP | Large-size | 4608–6480 | 100 | 100 000 | — |
| 2 | FEP | rungs 1–3 | Medium-size | 192–288 | 100 | 100 | 600 |
| 3 | FEP | rungs 4,5 | Medium-size | 192–288 | 1 | 100 | 600, 500 |
| 4 | FEP | rung 5 | Unit cell | 9–24 | 1 | 1 | 1000 |

TABLE S5. Supercells used in the direct-upsampling method. Here $\Delta F^{\mathrm{qh}\to\mathrm{MTP}}$ denotes the difference in Helmholtz energy with respect to the effective *harmonic* potential.

| | Small-size | | Medium-size | | Large-size | |
|---|---|---|---|---|---|---|
| | Supercell | Atoms | Supercell | Atoms | Supercell | Atoms |
| $\beta$-cristobalite | $1 \times 1 \times 1$ | 24 | $2 \times 2 \times 2$ | 192 | $6 \times 6 \times 6$ | 5184 |
| $\beta$-quartz | $2 \times 2 \times 2$ | 72 | $3 \times 3 \times 3$ | 243 | $8 \times 8 \times 8$ | 4608 |
| $P6_3/mmc$-tridymite | $2 \times 2 \times 1$ | 48 | $3 \times 3 \times 2$ | 216 | $9 \times 9 \times 6$ | 5832 |
| $C222_1$-tridymite | – | – | $3 \times 2 \times 2$ | 288 | $9 \times 5 \times 6$ | 6480 |
| Used for: | Low-MTP fitting | | High-MTP fitting, direct upsampling $\Delta F^{\mathrm{up}}$ | | Thermodynamic integration $\Delta F^{\mathrm{qh}\to\mathrm{MTP}}$ | |

TABLE S6. Low-converged DFT (low-DFT) parameters for *ab initio* MD at high temperatures. The low-DFT runs for the low-MTP fitting were performed with LDA. The LDA-based low-MTPs were used for the fitting of all high-MTPs.

| | Input | | | | | | | Output |
|---|---|---|---|---|---|---|---|---|
| | Supercell | Atoms | Cutoff[a] (eV) | $k$ points | Volume (ÅÅ³/atom) | Time step (fs) | Steps | Temperature (K) | Min. distance[c] (Å) |
| $\beta$-cristobalite | $1 \times 1 \times 1$ | 24 | 400 | $2 \times 2 \times 2$ | 14.29 ... 16.88, 4 $V$'s | 5 | 1000 | 1986[b] | 0.51 |
| $\beta$-quartz | $2 \times 2 \times 2$ | 72 | 400 | $2 \times 2 \times 2$ | 12.02 ... 14.39, 4 $V$'s | 5 | 1000 | 1986[b] | 1.30 |
| $P6_3/mmc$-tridymite | $2 \times 2 \times 1$ | 48 | 400 | $2 \times 2 \times 2$ | 14.73 ... 17.54, 16 $V$'s | 5 | 1000 | 1500 | 1.29 |

[a] The energy cutoff corresponds to the maximum `ENMAX`.
[b] This is the experimental melting temperature.
[c] "Min. distance" is the minimum distance between any two atoms during the MD.

TABLE S7. Snapshot-generation parameters with low-MTP MD in medium-size supercells at the experimental melting point of 1986 K.

| | Input | | | | | Output | |
|---|---|---|---|---|---|---|---|
| | Supercell | Atoms | Volume (Å³/atom) | Temperature (K) | Initial snapshots[a] | Selected snapshots[a] | Min. distance (Å) |
| **$\beta$-cristobalite** | | | | | | | |
| LDA | $2 \times 2 \times 2$ | 192 | 14.01 ... 15.55, 4 $V$'s $\times$ 3 $c/a$'s | 200, 400, 1986 | 2660 | 490 | 1.38 |
| PBE-D3(BJ) | $2 \times 2 \times 2$ | 192 | 14.0 ... 16.0, 8 $V$'s | 1986 | 240 | 236 | 1.37 |
| r²SCAN | $2 \times 2 \times 2$ | 192 | 14.0 ... 16.0, 8 $V$'s | 1986 | 231 | 171 | 1.38 |
| **$\beta$-quartz** | | | | | | | |
| LDA | $3 \times 3 \times 3$ | 243 | 12.02 ... 14.39, 8 $V$'s | 1986 | 240 | 204 | 1.39 |
| PBE-D3(BJ) | $3 \times 3 \times 3$ | 243 | 12.0 ... 14.0, 8 $V$'s | 1986 | 240 | 188 | 1.38 |
| r²SCAN | $3 \times 3 \times 3$ | 243 | 12.0 ... 14.0, 8 $V$'s | 1986 | 240 | 196 | 1.37 |
| **$P6_3/mmc$-tridymite** | | | | | | | |
| LDA | $3 \times 3 \times 2$ | 216 | 14.73 ... 17.54, 8 $V$'s | 1500 | 240 | 164 | 1.41 |
| PBE-D3(BJ) | $3 \times 3 \times 2$ | 216 | 14.73 ... 17.54, 8 $V$'s | 1500 | 240 | 164 | 1.39 |
| r²SCAN | $3 \times 3 \times 2$ | 216 | 14.73 ... 17.54, 8 $V$'s | 1500 | 240 | 162 | 1.38 |

[a] Initial snapshots are uncorrelated snapshots with a fixed time interval in between, and the selected snapshots are those selected for the training with an extrapolation grade threshold.

TABLE S8. Low-MTP fitting parameters and results. The low-DFT runs for the low-MTP fitting were performed with LDA. The LDA-based low-MTPs were used for the fitting of all high-MTPs.

| | Input | | | | Output | | |
|---|---|---|---|---|---|---|---|
| | $R_{min}$ (Å) | $R_{cut}$ (Å) | Level | No. of MTP parameters | Energy RMSE (meV/atom) | Force RMSE (eV/Å) | $\Delta F^{up}$ (meV/atom) |
| $\beta$-cristobalite | 2.0 | 6.0 | 12 | 128 | 5.4 | 1.686 | – |
| $\beta$-quartz | 1.30 | 6.0 | 10 | 115 | 2.7 | 0.165 | – |
| $P6_3/mmc$-tridymite | 1.28 | 6.0 | 10 | 115 | 2.2 | 0.114 | – |

TABLE S9. High-MTP fitting parameters and results. The minimum distance and cutoff radius of the MTP are denoted as $R_{min}$ and $R_{cut}$, respectively. The $\Delta F^{up}$ column refers to the Helmholtz energy difference between the MTP and the respective exchange-correlation functional. This difference is evaluated at the $(V,T)$ grid point closest to $T_{q\text{-}c}^{exp}$ and the corresponding volume.

| | Input | | | | Output | | |
|---|---|---|---|---|---|---|---|
| | $R_{min}$ (Å) | $R_{cut}$ (Å) | Level | No. of MTP parameters | Energy RMSE (meV/atom) | Force RMSE (eV/Å) | $\Delta F^{up}$ (meV/atom) |
| $\beta$-cristobalite | | | | | | | |
| LDA | 2.0 | 6.0 | 16 | 223 | 0.52 | 0.066 | 0.07 |
| PBE-D3(BJ) | 1.35 | 6.0 | 16 | 223 | 0.97 | 0.087 | 0.18 |
| r²SCAN | 1.37 | 6.0 | 16 | 223 | 1.06 | 0.093 | 0.12 |
| $\beta$-quartz | | | | | | | |
| LDA | 1.37 | 6.0 | 18 | 326 | 0.73 | 0.091 | 0.09 |
| PBE-D3(BJ) | 1.38 | 6.0 | 16 | 223 | 1.06 | 0.109 | 0.10 |
| r²SCAN | 1.36 | 6.0 | 18 | 326 | 1.08 | 0.108 | 0.11 |
| $P6_3/mmc$-tridymite | | | | | | | |
| LDA | 1.39 | 6.0 | 16 | 223 | 0.60 | 0.086 | 0.13 |
| PBE-D3(BJ) | 1.38 | 6.0 | 16 | 223 | 1.02 | 0.097 | 0.13 |
| r²SCAN | 1.38 | 6.0 | 16 | 223 | 1.12 | 0.104 | 0.12 |

TABLE S10. Parameters for fitting effective harmonic potentials to high-DFT forces from high-MTP MD snapshots. Only LDA high-MTPs were used, and LDA harmonic potentials were fitted. The latter were then used for all functionals.

| | Input | | | | | | | | Output | |
|---|---|---|---|---|---|---|---|---|---|---|
| | Fitting | | Extended dyn. mat. | | | | | | | |
| | Supercell | Atoms | Supercell | Atoms | Volume (Å³/atom) | Temperature (K) | $R_{cut}$ (Å) | $N_{param}$[a] | $R^2$ | Force RMSE (eV/Å) |
| $\beta$-cristobalite | 2×2×2 | 192 | 6×6×6 | 5184 | 15.71 | 1500 | 6.6 | 60 | 0.72 | 1.09 |
| $\beta$-quartz | 3×3×3 | 243 | 8×8×8 | 4608 | 13.57 | 1000 | 6.5 | 312 | 0.86 | 0.61 |
| $P6_3/mmc$-tridymite | 3×3×2 | 216 | 9×9×6 | 5832 | 16.09 | 1500 | 6.6 | 206 | 0.77 | 0.98 |
| $C222_1$-tridymite [b] | 3×2×2 | 288 | 9×5×6 | 6480 | 16.00 | 1986 | 4.5 | 295 | 0.71 | 1.23 |

[a] $N_{param}$ is the number of fitting parameters.
[b] The $C222_1$-tridymite effective harmonic potential was fitted to PBE-D3(BJ).

TABLE S11. High-DFT parameters in medium-size supercells. The settings are the same for all rung 1–3 functionals.

| | Supercell | Atoms | Cutoff[a] (eV) | $k$ points | kp·atoms |
|---|---|---|---|---|---|
| $\beta$-cristobalite | 2×2×2 | 192 | 600 | 1×1×1 | 192 |
| $\beta$-quartz | 3×3×3 | 243 | 600 | 1×1×1 | 243 |
| $P6_3/mmc$-tridymite | 3×3×2 | 216 | 600 | 1×1×1 | 216 |
| $C222_1$-tridymite | 3×2×2 | 288 | 600 | 1×1×1 | 288 |

[a] The energy cutoff corresponds to the maximum 1.5 × ENMAX in the VASP POTCAR.

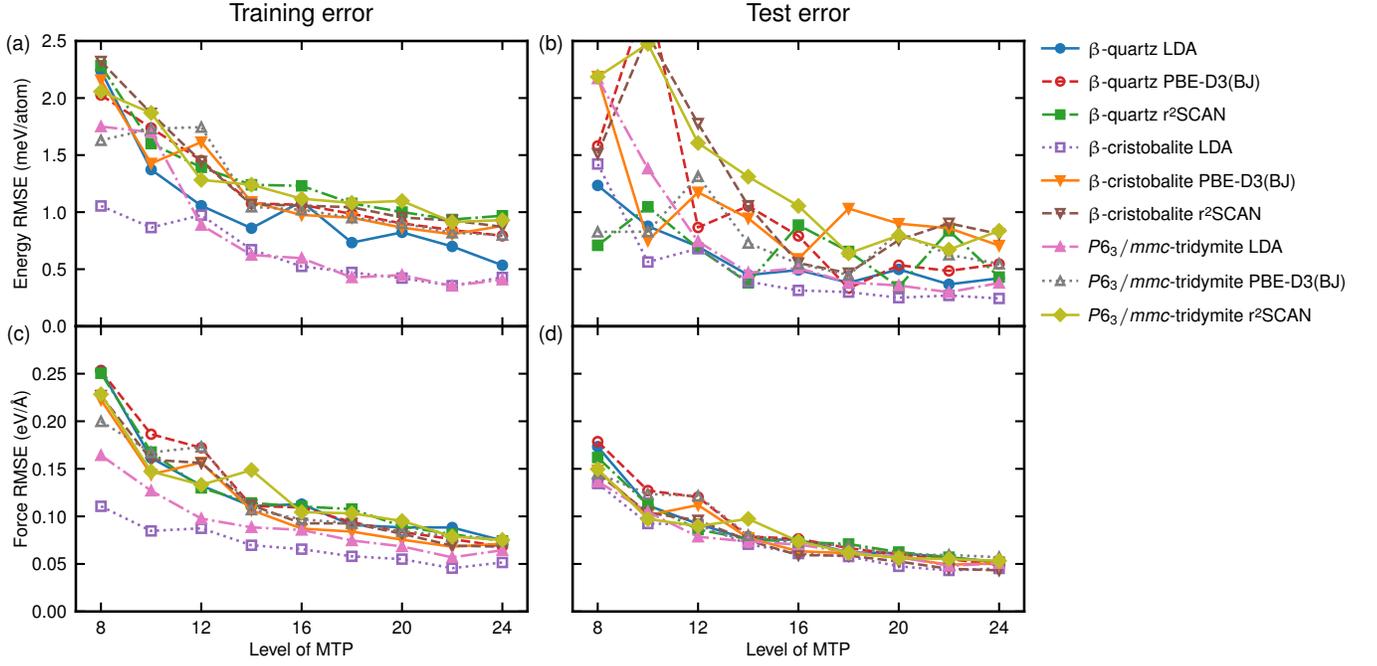

FIG. S8. Training and test errors of the high-MTPs. The training root-mean-squared errors (RMSEs) of the (a) energy per atom and (c) force components. The MTPs are trained in medium-size supercells to LDA, PBE-D3(BJ), and r²SCAN DFT energies and forces of configurations at high temperatures in the relevant $V$ range. Test RMSEs of the (b) energy per atom and (d) force components corresponding to configurations of the upsampling at the $T$ grid point of transition from $\beta$-quartz to $\beta$-cristobalite $T_{q\text{-}c}^{\text{exp}} = 1137$ K and the equilibrium $V$.

TABLE S12. Parameters for the anharmonic calculations; thermodynamic integration (TI) from effective harmonic potential to high-MTP and upsampling (or FEP) from high-MTP to high-DFT. For the other functionals in rungs 1–3 only upsampling was performed from PBE-D3(BJ) and r²SCAN with the same cells and on the same grids as listed here.

| | TI | | Upsampling/FEP | | | | |
|---|---|---|---|---|---|---|---|
| | Supercell | Atoms | Supercell | Atoms | Snapshots[a] | Volume (Å³/atom) | Temperature (K) |
| $\beta$-cristobalite | | | | | | | |
| LDA | $6 \times 6 \times 6$ | 5184 | $2 \times 2 \times 2$ | 192 | 80 | 14.21 ... 15.93, 25 $V$'s | 1986/7×{3, 4, 5, 6, 7, 7.5, 8, 9, 10} |
| PBE-D3(BJ) | $6 \times 6 \times 6$ | 5184 | $2 \times 2 \times 2$ | 192 | 80 | 14.00 ... 16.00, 11 $V$'s | 1986/7×{3, 4, 5, 6, 7, 7.5, 8, 9} |
| r²SCAN | $6 \times 6 \times 6$ | 5184 | $2 \times 2 \times 2$ | 192 | 80 | 14.00 ... 16.00, 11 $V$'s | 1986/7×{3, 4, 5, 6, 7, 7.5, 8, 9} |
| $\beta$-quartz | | | | | | | |
| LDA | $8 \times 8 \times 8$ | 4608 | $3 \times 3 \times 3$ | 243 | 80 | 12.21 ... 14.39, 12 $V$'s | 1986/7×{3, 4, 5, 6, 7, 7.5, 8, 9, 10} |
| PBE-D3(BJ) | $8 \times 8 \times 8$ | 4608 | $3 \times 3 \times 3$ | 243 | 80 | 13.00 ... 14.00, 11 $V$'s | 1986/7×{3, 4, 5, 6, 7, 7.5, 8, 9} |
| r²SCAN | $8 \times 8 \times 8$ | 4608 | $3 \times 3 \times 3$ | 243 | 80 | 12.00 ... 14.00, 11 $V$'s | 1986/7×{3, 4, 5, 6, 7, 7.5, 8, 9} |
| $P6_3/mmc$-tridymite | | | | | | | |
| LDA | $9 \times 9 \times 6$ | 5832 | $3 \times 3 \times 2$ | 216 | 80 | 14.15 ... 16.09, 20 $V$'s | 1986/7×{3, 4, 5, 6, 7, 7.5, 8, 9, 10} |
| PBE-D3(BJ) | $9 \times 9 \times 6$ | 5832 | $3 \times 3 \times 2$ | 216 | 80 | 14.00 ... 16.00, 11 $V$'s | 1986/7×{3, 4, 5, 6, 7, 7.5, 8, 9} |
| r²SCAN | $9 \times 9 \times 6$ | 5832 | $3 \times 3 \times 2$ | 216 | 80 | 14.00 ... 16.00, 11 $V$'s | 1986/7×{3, 4, 5, 6, 7, 7.5, 8, 9} |
| $C222_1$-tridymite | | | | | | | |
| LDA | $9 \times 5 \times 6$ | 6480 | $3 \times 2 \times 2$ | 288 | 80 | 14.00 ... 16.00, 11 $V$'s | 1986/7×{3, 4, 5, 6, 7, 7.5, 8, 9} |
| PBE-D3(BJ) | $9 \times 5 \times 6$ | 6480 | $3 \times 2 \times 2$ | 288 | 80 | 14.00 ... 16.00, 11 $V$'s | 1986/7×{3, 4, 5, 6, 7, 7.5, 8, 9} |
| r²SCAN | $9 \times 5 \times 6$ | 6480 | $3 \times 2 \times 2$ | 288 | 80 | 14.00 ... 16.00, 11 $V$'s | 1986/7×{3, 4, 5, 6, 7, 7.5, 8, 9} |

[a] The high-DFT sampling of snapshots is finished when the estimated deviation becomes less than a target value at each $(V, T)$ grid point. The number of sampled snapshots is variable, and the maximum numbers of snapshots are listed.

TABLE S13. Basis-set elements used for the parametrization of the anharmonic Helmholtz energy contribution $F^{\text{ah}}$.

| Order | 0 | 1 | 2 | 3 | 4 |
|---|---|---|---|---|---|
| Basis element[a] | 1 | $V, T$ | $V^2, VT, T^2$ | $V^3, V^2T, VT^2$ | $V^3T, V^2T^2$ |

[a] For $\beta$-quartz with the vdW-DF-cx functional, the $V^3T$-term was excluded.

## B. Systematic tests for convergence and accuracy

In Fig. S9, we show convergence tests for cristobalite for the supercell size. The behavior is similar also for tridymite, and for production runs a size of around 5000 atoms was chosen for all phases.

In Fig. S10, we show the convergence with respect to the plane-wave cutoff for the upsampling to r²SCAN. In (a) and (b), we show the energies of five randomly chosen snapshots for $\beta$-quartz and $\beta$-cristobalite, respectively. By showing the energy offset of each snapshot from the average energy (at each energy cutoff), the effect on second- and higher-order terms in the free-energy perturbation is isolated. (The dependence of the absolute energies on the cutoff is, of course, much more severe.) In (c), we in turn show the dependence of the energy difference between $\beta$-quartz and $\beta$-cristobalite, calculated from their average snapshot energies. The energy difference between the phases converges slightly slower than the relative energies of the snapshots within each phase. Together, these results ensure that our final values are converged both in terms of upsampling for each phase and in terms of phase stability. Further convergence tests for the $k$-point density indicate that already around 200 kp · atoms provides very well converged results. Tests conducted up to around 25 000 kp · atoms for upsampling to r²SCAN, show maximum energy differences of only 0.04 meV/atom.

Convergence of the order of the polynomials used for fitting and the $(V, T)$ grid can be seen in Fig. S11 and Fig. S12, respectively. Figure S11 shows the volume dependence of the Helmholtz energy at the $\beta$-quartz–$\beta$-cristobalite transition temperature, and the temperature dependence of the volume and bulk modulus at ambient pressure, for different polynomial basis sets. The 2nd-order basis is not enough to capture the volume dependence of the Helmholtz energy surface. A 3rd-order polynomial basis, on the other hand, captures the temperature dependence well, and only a small change is seen compared to the 4th-order basis in Helmholtz energy and equi-

librium volume. For the $(V, T)$-grid convergence, at least the $11 \times 8$ grid is necessary for a sufficiently converged Helmholtz energy. For the bulk modulus, there is a clear difference between 3rd and 4th order polynomial basis, especially for $\beta$-quartz. For the bulk modulus of $\beta$-quartz there is also a higher sensitivity to the $(V, T)$-grid density. Since the main focus of this work is phase stabilities and transition properties, we have used the 4th-order basis as presented in Table S13 and at least an $11 \times 8$ $(V, T)$ grid for the production calculations.

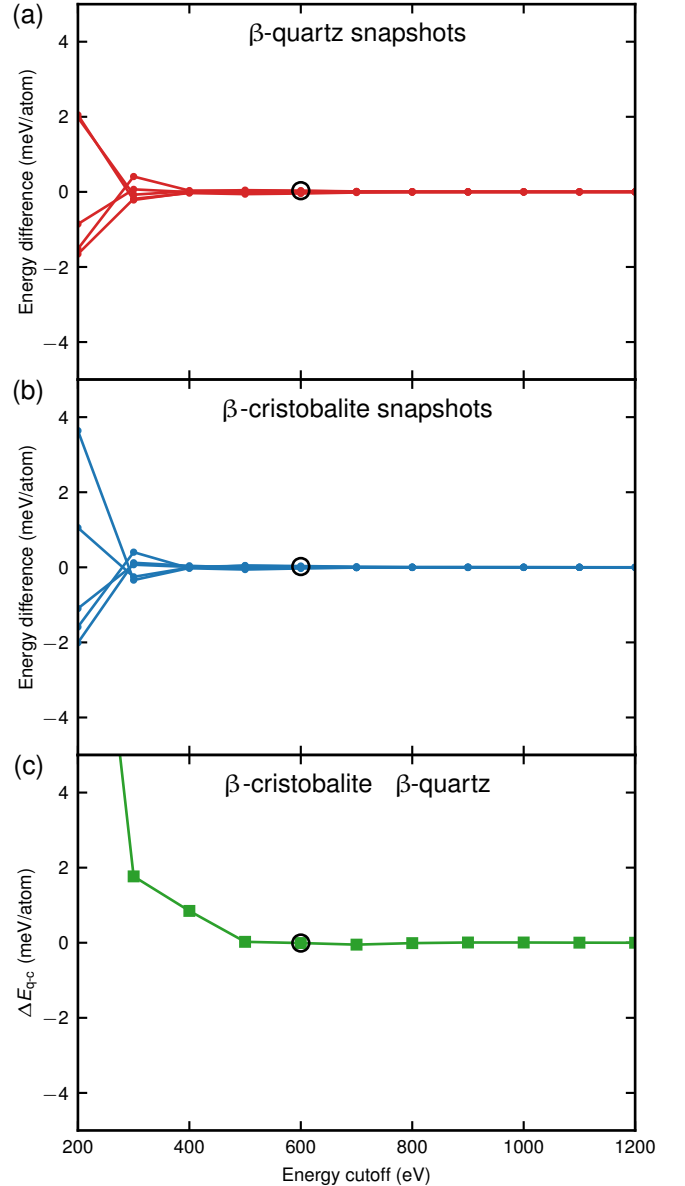

FIG. S10. r²SCAN energy convergence with respect to the plane-wave cutoff (ENCUT), between snapshots within the same phase [(a) and (b)] and for the snapshot-average difference between $\beta$-quartz and $\beta$-cristobalite $\Delta E_{\text{q-c}}$ in (c). In (a) and (b), the energies are referenced with respect to the average snapshot energy for each cutoff. All curves are shifted with respect to their last value. The values used for production calculations are marked by a circle.

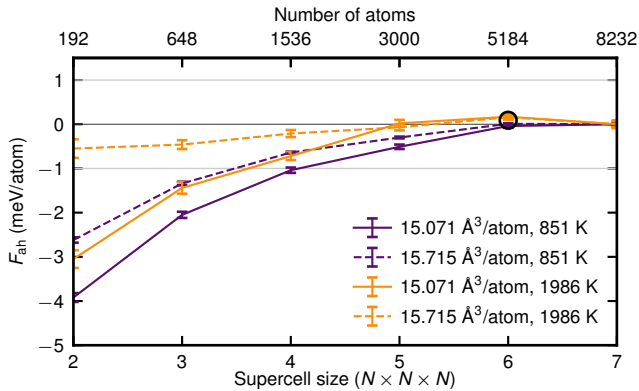

FIG. S9. Convergence of the anharmonic Helmholtz energy $F^{\text{ah}}$ with respect to supercell size for $\beta$-cristobalite computed with the LDA MTP. The error bars indicate the statistical errors within a 95% confidence interval, and the black circle marks the size used for TI in the production runs.

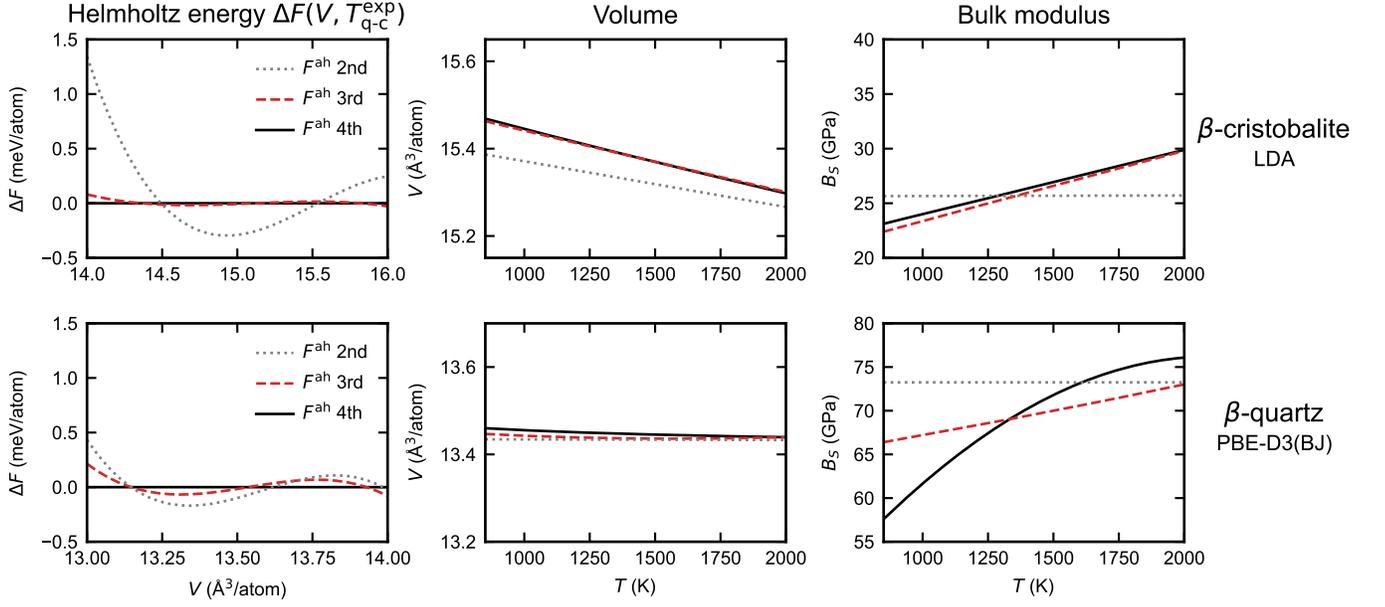

FIG. S11. Dependence of thermodynamic properties on the order of the polynomial basis used in the parametrization of the anharmonic Helmholtz energy $F^{\text{ah}}(V,T)$, for $\beta$-cristobalite with LDA and $\beta$-quartz with PBE-D3(BJ), sampled on a $11V \times 8T$ grid. Here, $\Delta F(V, T_{\text{q-c}}^{\text{exp}})$ denotes the difference in Helmholtz energy from the polynomial with the highest order 4, at 1137 K. In addition, the volume and the adiabatic bulk modulus $B_S$ are shown.

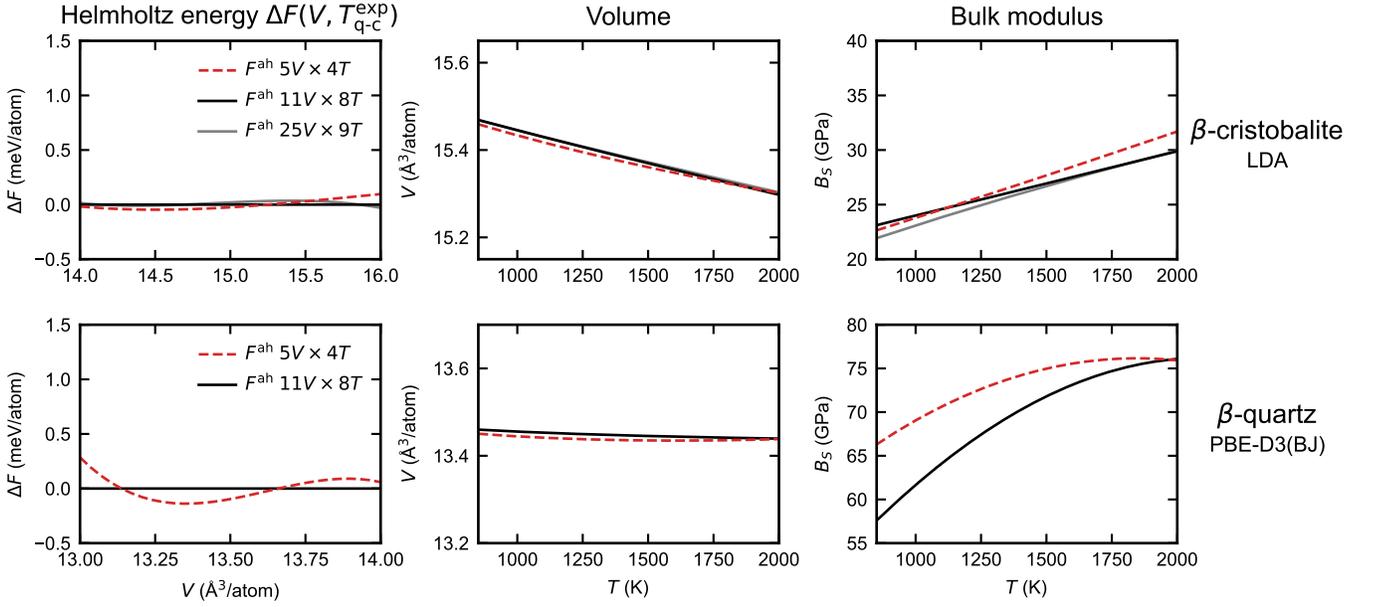

FIG. S12. Dependence of thermodynamic properties on the grid density for the anharmonic Helmholtz energy $F^{\text{ah}}(V,T)$. Here, $\Delta F(V, T_{\text{q-c}}^{\text{exp}})$ denotes the difference in Helmholtz energy from the $11V \times 8T$ grid at 1137 K, and $B_S$ the adiabatic bulk modulus.

## C. RPA details and convergence tests

Due to the significant computational requirements of RPA calculations, a special procedure was devised to obtain the RPA results presented in Table S2. To compute the RPA up-sampling energy $\Delta F^{up}$, around 50 snapshots are needed to provide a good statistical convergence. The size of the supercell for the upsampling is around 200 atoms, specifically 243 atoms for $\beta$-quartz and 192 atoms for $\beta$-cristobalite. In order for such calculations to be computationally feasible, we compute in the first step of the procedure $\Delta F^{up}_{normal}$ with a set of normally converged parameters. The requirement on these parameters is to provide converged second-order and higher-order terms of the free-energy perturbation. In the second step, we add a first-order correction calculated with a smaller snapshot (9/24 atoms for $\beta$-quartz/$\beta$-cristobalite). Using a smaller snapshot allows us to use highly converged computational parameters that ensure the highest precision, also with respect to the relative stability between the phases. This first-order correction is strictly valid when the energy of the smaller snapshot is converged with respect to the snapshot average energy. Ultimately, the upsampling Helmholtz energy is obtained as

$$\Delta F^{up} = \Delta F^{up}_{normal} + E_{high} - E_{normal}, \quad (S6)$$

where $E_{high}$ is the highly converged energy per atom of the small snapshot and $E_{normal}$ is the energy per atom of the small snapshot, scaled to 243 or 192 atoms and with identical parameters to the upsampling. The energies for these steps are detailed in Table S14, Table S15, and Table S16. The computational parameters used in each step can be found in Table S17.

In order to guarantee the validity and precision of the approach, extensive convergence tests were performed. In Fig. S13, we show the dependence of absolute energies on the number of time/frequency points, at an offset from the highest value. There is, in principle, no dependence on the number of time/frequency points, even on the absolute scale. Only for higher cutoffs, above 1000 eV, could we notice a small time/frequency dependence. The convergence results for these high cutoffs were obtained with an increased number of points, 16 points. The $k$-point dependence is also weak, on the relative scale. In Fig. S14, we show the dependence of the energy difference between two snapshots within each phase, respectively, as well as the energy difference between $\beta$-quartz and $\beta$-cristobalite. This applies both to the RPA correlation energy and the exact exchange energy when a cutoff for the Coulomb kernel is used for the Fock and exchange energy evaluation. (Here, we used HFRCUT = −1 for an automated cutoff in VASP.) Still, we applied the first-order correction to the exact exchange energy and finally used 1125/1536 kp · atoms for $\beta$-quartz/$\beta$-cristobalite. For $\Delta F^{up}_{normal}$ only the $\Gamma$ point was sampled, i.e., 243/192 kp · atoms for $\beta$-quartz/$\beta$-cristobalite.

The energy-cutoff convergence is more severe, as seen in Fig. S15. In the left column, the convergence of the exact exchange energy is shown, and in the right column, that of the RPA correlation energy. In a similar way as in Fig. S10, we show the convergence of $\beta$-quartz snapshots in (a) and (b), the convergence of $\beta$-cristobalite snapshots in (c) and (d), and finally the energy difference between $\beta$-quartz and $\beta$-cristobalite in (e) and (f). Additionally, a thicker line is shown in (a)–(d), indicating the convergence of the small snapshot with respect to the snapshot average. Using the converged cutoff marked by circles guarantees that the first-order correction is valid and yields precise results. The phase energy difference in (e) and (f) was calculated with the small snapshot for each phase. Using the small snapshot, we could also investigate the effect of the Si core polarization—in practice by performing RPA calculations with a PAW potential including the Si $2s$ and $2p$ electrons in the valence shell.

In Table S17, the RPA computational parameters are presented. For smooth convergence and accurate results, PRECFOCK=Normal for both exact exchange and RPA calculations was important. PRECFOCK=Accurate gave only very small differences in energy in the investigated system. The computational load increased by around 25% when using PRECFOCK=Normal compared to PRECFOCK=Fast. On the other hand, using PRECFOCK=Normal allowed for using at least 50 eV lower ENCUT (500 eV instead of 550 eV), thereby almost fully compensating for using PRECFOCK=Normal in terms of computation time. Additionally, PRECFOCK=Normal was important to obtain an agreement between the energies per atom of the scaled and the small cells, when computed with corresponding settings.

For the calculations of the exact exchange energy, HFRCUT is in this case important for a reasonable $k$-point convergence.

## D. Expansion of free-energy perturbation

The convergence behavior of the upsampling is quantified by the second-order term of the expansion of the thermodynamic free-energy perturbation [41] formula in Eq. (4) in the main text:

$$\Delta F^{up}(V, T) = -k_B T \ln \left\langle \exp\left(-\frac{\Delta E}{k_B T}\right) \right\rangle_{ML}$$
$$= \omega_1 - \omega_2/2k_B T + O\left((1/k_B T)^2\right), \quad (S7)$$

where $\omega_1 = \langle \Delta E \rangle_{ML}$, $\omega_2 = \left\langle (\Delta E)^2 \right\rangle_{ML} - \langle \Delta E \rangle^2_{ML}$, and $\Delta E = E - E^{ML}$, with the energies $E$ and $E^{ML}$ calculated with the functional of interest and the ML potential, respectively. The second term of the second line of Eq. (S7),

$$e_2 = -\frac{\omega_2}{2k_B T} = -\frac{1}{2k_B T}\left[\frac{1}{N_s}\sum_{i=1}^{N_s}\Delta E_i^2 - \left(\frac{1}{N_s}\sum_{i=1}^{N_s}\Delta E_i\right)^2\right], \quad (S8)$$

is called the second-order term. The right-hand side of Eq. (S8) expresses the second-order term in the notation of Sec. S1 A.

TABLE S14. Upsampling results from PBE-D3(BJ) MTP snapshots to RPA with "normal" settings, i.e., a plane-wave energy cutoff of 500 eV. The statistical error of $\Delta F_{\text{normal}}^{\text{up}}$ is given within a 95% confidence interval and is estimated through the delta method [40].

|  | $\Delta F_{\text{normal}}^{\text{up}}$ (meV/atom) | Second-order term (meV/atom) | Number of snapshots |
|---|---|---|---|
| $\beta$-quartz | $-16\,206.57 \pm 0.18$ | $-0.45$ | 51 |
| $\beta$-cristobalite | $-16\,211.23 \pm 0.23$ | $-0.62$ | 51 |

TABLE S15. RPA first-order correction $E_{\text{high}} - E_{\text{normal}}$ (from 500 to 1000 eV plane-wave energy cutoff) and the separate terms $E_{\text{normal}}$ and $E_{\text{high}}$ for $\beta$-quartz and $\beta$-cristobalite. The contributions from exchange and correlation are given together with the total value for each phase. The unit is meV/atom.

|  | $E_{\text{normal}}$ | $E_{\text{high}}$ | $E_{\text{high}} - E_{\text{normal}}$ |
|---|---|---|---|
| $\underline{\beta\text{-quartz}}$ | | | |
| Exchange | $-15\,979.41$ | $-15\,978.94$ | $0.47$ |
| Correlation | $-8170.63$ | $-8263.35$ | $-92.71$ |
| Total | $-24\,150.05$ | $-24\,242.28$ | $-92.24$ |
| $\underline{\beta\text{-cristobalite}}$ | | | |
| Exchange | $-15\,951.44$ | $-15\,951.00$ | $0.44$ |
| Correlation | $-8171.26$ | $-8263.32$ | $-92.05$ |
| Total | $-24\,122.70$ | $-24\,214.32$ | $-91.62$ |

TABLE S16. Final RPA results. The statistical error is given within a 95% confidence interval and is estimated through the delta method [40]. The unit is meV/atom.

|  | $\beta$-quartz | $\beta$-cristobalite | Difference $\Delta F_{\text{q-c}}$ |
|---|---|---|---|
| $\Delta F_{\text{normal}}^{\text{up}}$ | $-16\,206.57 \pm 0.18$ | $-16\,211.23 \pm 0.23$ | $-4.65 \pm 0.29$ |
| $E_{\text{high}} - E_{\text{normal}}$ | $-92.24$ | $-91.62$ | $0.62$ |
| Sum $\Delta F^{\text{up}}$ | $-16\,298.81$ | $-16\,302.84$ | $-4.04$ |
| $F^{\text{ML}}(V, T; \text{PBE-D3(BJ)})$ | — | — | $2.07$ |
| Core polarization shift (Si_sv_GW) | — | — | $1.71$ |
| RPA (without core polarization) | — | — | $-1.97$ |
| RPA | — | — | $-0.26$ |

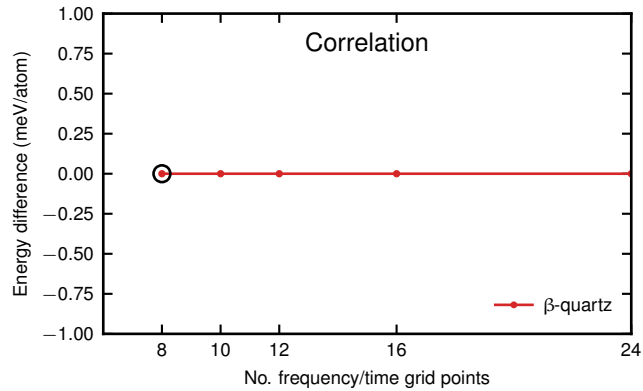

FIG. S13. RPA energy convergence with respect to the number of frequency and time grid points (`NOMEGA`), between two snapshots within $\beta$-quartz. The value used for production calculations is marked by a circle. The plane-wave energy cutoff was 450 eV.

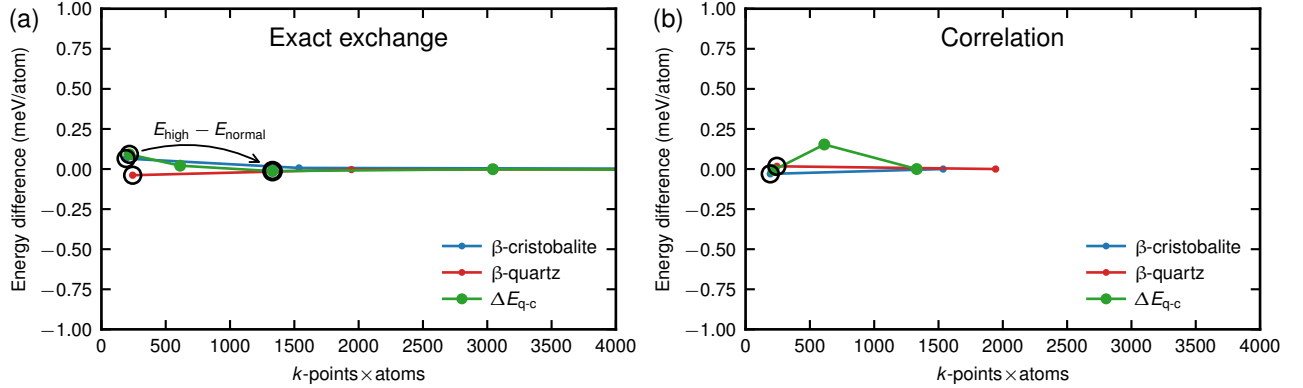

FIG. S14. RPA energy convergence with respect to $k$ points, between two snapshots within the same phase (red and blue) and for the energy difference between one $\beta$-quartz and one $\beta$-cristobalite snapshot $\Delta E_{\text{q-c}}$ (green), for (a) the exact exchange energy and (b) the correlation energy. The values used for production calculations are marked by a circle. The plane-wave energy cutoff was 400 eV for the exact exchange energy and 350 eV for the correlation, except for the green lines where 650 eV was used. Note that the energy scale differs from Fig. S15 by a factor of 1/5.

TABLE S17. Computational parameters used for the RPA calculations. The "High" settings were used for calculating $E_{\text{high}}$ and the "Normal" settings for $E_{\text{normal}}$ and $\Delta F_{\text{normal}}^{\text{up}}$.

| | Normal convergence (between snapshots) | | High convergence (between phases) | |
| --- | --- | --- | --- | --- |
| | Exact exchange | Correlation | Exact exchange | Correlation |
| PREC | Accurate | Accurate | Accurate | Accurate |
| PRECFOCK[a] | Normal | Normal | Normal | Normal |
| LMAXFOCKAE | 4 | 4 | 4 | 4 |
| NOMEGA | — | 8 | — | 8 |
| ENCUT | 500 | 500 | 1000 | 1000 |
| ENCUTGW | 333[b] | 333[b] | 667[b] | 667[b] |
| ENCUTGWSOFT | 267[c] | 267[c] | 533[c] | 533[c] |
| HFRCUT | −1 | — | −1 | — |
| $k$-points×atoms $\beta$-quartz | 243 | 243 | 1125[d] | 243 |
| $k$-points×atoms $\beta$-cristobalite | 192 | 192 | 1536[d] | 192 |

[a] PRECFOCK affects the FFT grid densities and at least the Normal setting was required.
[b] This is the default VASP value (2/3*ENCUT).
[c] This is the default VASP value (0.8*ENCUTGW).

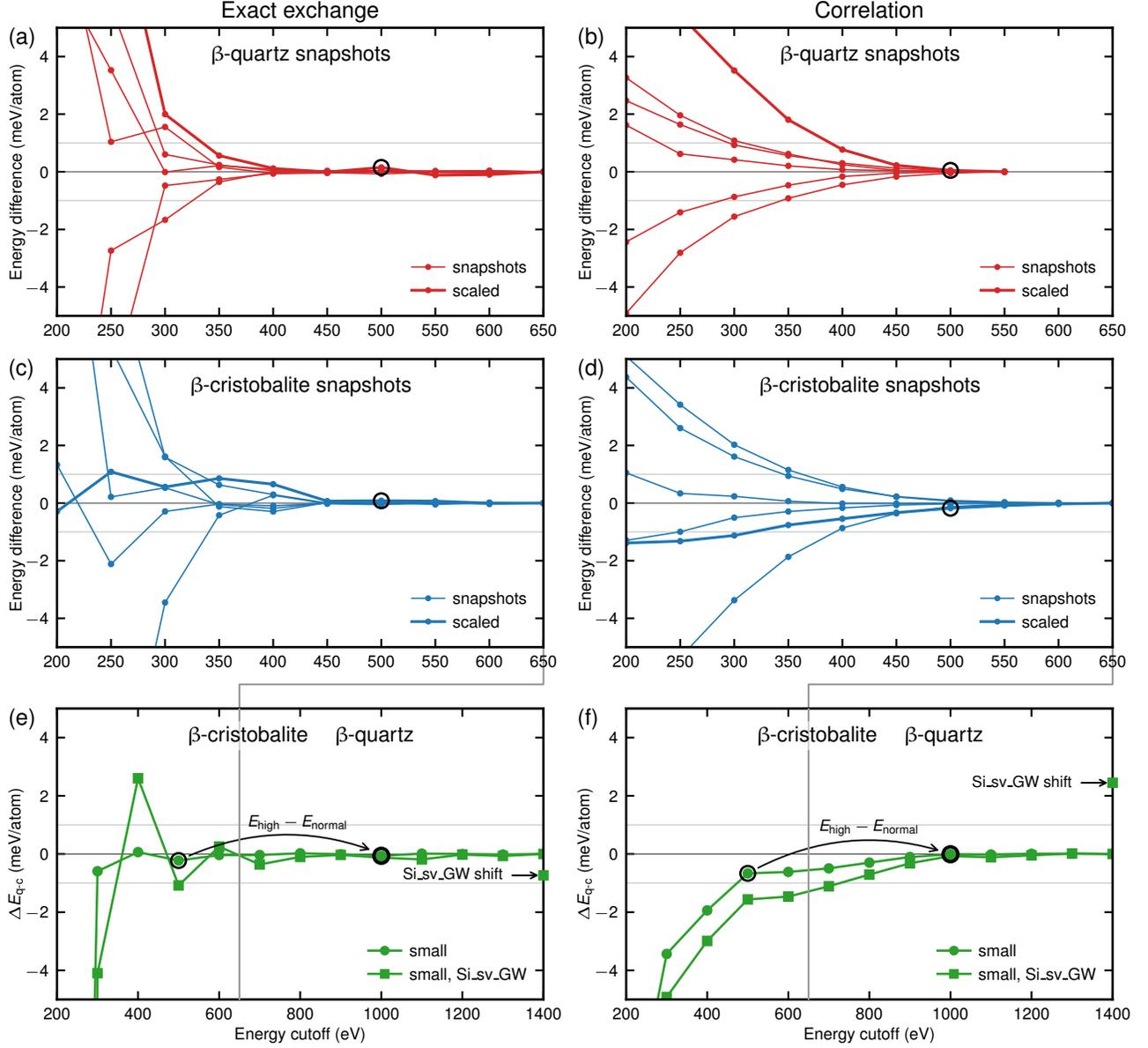

FIG. S15. RPA energy convergence with respect to the plane-wave cutoff (`ENCUT`), between snapshots within the same phase [(a)–(d)] and for the energy difference between $\beta$-quartz and $\beta$-cristobalite small cells $\Delta E_{\text{q-c}}$ [(e) and (f)]. In (a)–(d), the energies are referenced with respect to the average snapshot energy for each cutoff. All curves are shifted with respect to their last value. Further, the shifts resulting from core polarization (Si semi-core states included) are marked with additional points. The values used for production calculations are marked by a circle.


\fi


\end{document}